\documentclass[12pt, a4paper, twoside]{article}

\usepackage{latexsym}
\usepackage{epsfig} %\includegraphics
\usepackage{amsmath}
\usepackage{amssymb}
\usepackage{theorem}
\usepackage{apa}

\newtheorem{thm}{Theorem}[section] % Nummereringen af sætninger følger 
\newtheorem{prop}[thm]{Proposition}

\DeclareMathOperator{\tr}{\tau}
\DeclareMathOperator{\var}{Var}

\DeclareMathOperator{\corr}{Cor}

\title{A new approach to multi-modal diffusions\\ with applications to protein folding}

\author{Julie Lyng Forman${}^{1,\star}$, Michael S{\o}rensen${}^{2}$ \bigskip
  \\
  ${}^{1}${\small Department of Biostatistics},
  {\small University of Copenhagen}\\ {\small \O ster Farimagsgade 5},
  {\small DK-1014 Copenhagen K}
  \medskip
  \\
  ${}^{2}${\small Department of Mathematical Sciences,}
  {\small University of Copenhagen}\\
   {\small Universitetsparken 5,}
  {\small DK-2100 Copenhagen {\O}}
  \medskip
  \\
  {\small $^\star$Email:} {\small {\tt jufo@biostat.ku.dk}}}

\date{}

\begin{document}
\maketitle

\begin{center}
{\bf Summary}
\end{center}
This article demonstrates that flexible and statistically tractable
multi-modal diffusion  models can be attained by transformation of
simple well-known diffusion models such as the Ornstein-Uhlenbeck
model, or more generally a Pearson diffusion. The transformed
diffusion inherits many properties of the underlying simple diffusion
including its mixing rates and distributions of first passage
times. Likelihood inference and martingale estimating functions are
considered in the case of a discretely observed bimodal diffusion.
It is further demonstrated that model parameters can be identified and
estimated when the diffusion is observed with additional measurement
error. The new approach is applied to molecular dynamics data in form 
of a reaction coordinate of the small Trp-zipper protein, for which the 
folding and unfolding rates are estimated. The new models provide a better 
fit to this type of protein folding data than previous models because the 
diffusion coefficient is state-dependent. 
\medskip


\section{Introduction}
In this article we propose a new class of stationary stochastic
differential equation models that have multi-modal invariant
distributions. These models are useful for modeling dynamical systems
that switch randomly between two or more regimes. As an example, we
consider molecular dynamics data in form of a protein reaction coordinate 
with two regimes corresponding to the folded and unfolded state of the protein, respectively.
Essentially protein folding happens in a high-dimensional space, but a 
remarkable consequence of the
energy landscape theory is that folding is essentially a low dimensional process as it happens 
down a folding funnel. It has been suggested that folding can be accurately captured by one or 
a few suitably chosen protein reaction coordinates (that is, univariate characteristics of the 
protein) with diffusive dynamics along the folding funnel, see \cite{sow:96}, \cite{das:06},
and the references therein. Note, however, that applications of bimodal diffusions 
are not limited to the study of molecular dynamics. Other applications of bimodal diffusion 
are as models of the global climate where the two regimes could be a cold and a hot climate 
as in \cite{imkeller}, and as 
financial models of e.g.\ interest rates subject to changes in the underlying financial and 
economic mechanisms as in \cite{yas:96}.

Traditionally bimodal diffusion processes have been constructed by a stochastic differential 
equation with additive noise for a process moving in a double-well potential, i.e. a stochastic 
differential equation of the form
\begin{equation}
\label{e:dwpdif}
dY_t = - V'(Y_t)dt + \sigma^2dB_t,
\end{equation}
where $\{B_t\}$ is a Wiener process and $V$ is a potential with two valleys. Under the condition 
that $V(y)$ goes to infinity at the boundaries of the state space and that the function 
$h(y)=\exp\{-2V(y)/\sigma^2\}$ is integrable on the state space, $\{Y_t\}$ is ergodic with invariant 
density proportional to $h(y)$. An often studied example is given by the potential 
$V(y)=\theta y^2 (y^2-2)$ with $\theta >0$, for which the drift is 
$-4 \theta (y^3 - y) = - 4 \theta y (y+1)(y-1)$. This simple potential has wells of the same depths 
at 1 and -1 and is symmetric around a separating potential barrier at 0. The related diffusion is 
ergodic with invariant density proportional to $\exp\{-2\theta (y^4 - 2 y^2)\}$. It is easy to generalize 
this model to models with wells at other points that need not be symmetric around the separating potential 
barrier. The double well potential models are the state of the art in the analysis of protein 
reaction coordinates. That is, constant diffusion is usually assumed and used in the estimation 
of the protein folding rates, see e.g.\ \cite{sow:96}. 
A more complex model of molecular dynamics was presented in \cite{stuart} who 
used a partially observed hypoelliptical diffusion to model the dihedral angle 
in a butane molecule. Still this model assumes a constant diffusion coefficient which may conflict 
with that of the data.
More recently evidence of non-constant diffusion in protein reaction coordinates have been 
reported in several articles. \cite{bh:10} discusses these findings 
and their implications for the assessment of protein folding rates.

Our new class of bimodal diffusions is obtained by applying particular transformations to
simple well-known diffusions such as the Ornstein-Uhlenbeck process or a general Pearson 
diffusion; see \cite{fs:08}. This leads to diffusion models
with nonlinear drift and non-constant diffusion coefficients that are still highly tractable both
from a statistical and a computational point of view. A major point of this article is that many 
properties of diffusions are preserved by transformation. These include 
stationarity, mixing properties, and distributions of first passage times. 
Also the eigenvalues of the infinitesimal generator of the diffusion
are preserved by the transformation, and eigenfunctions transform in a
straightforward way. This facilitates efficient approximate likelihood 
inference by means of e.g.\ the explicit martingale estimating
functions proposed by \cite{ks:99}. In the rare
cases where the likelihood function of a diffusion is explicitly known, this is also the case for any
of its transformations. Similarly to the double well potential models, our new diffusion models 
allow for great flexibility in the modeling of the invariant marginal distribution. Thus, the new bimodal 
diffusion models provide a useful extension of the class of bimodal
diffusion models.

An alternative to the stochastic differential equation approach is to model each regime 
separately and to let the shifts between the models be determined by an underlying 
finite-state process such as a hidden Markov model or Markov state model. These models are 
widely employed as models of protein folding, see \cite{prinz:11} for a review, although it 
is recognized that the models are inadequate in describing the more gradual transition between 
states which is the de facto behaviour of many proteins. Latent Markov state models are also 
very popular in financial and econometric applications, see \cite{lr:09} for an overview. 
However, a latent state model is too complex and difficult to interpret if what is observed is not two 
different dynamical systems, but is really the same dynamical system that just has the property that it 
can be in two different regimes. A multi-modal diffusion has local attractions points corresponding to 
its regimes and moves between them in a continuous and random way. Apart from the conceptual advantage 
and the simpler model, other advantages of a multi-modal diffusion over two separate models is that the stationary
marginal density in a succinct way contains important information about the regimes: the relative size 
of the modes reflect the time spent in each mode, and the peakedness and broadness of the modes reflect 
the volatility of the regimes. Moreover, explicit formulae for mean passage times allows for precise 
calculation of the time spent in each regime and the probability of
switching from one regime to another.

The article is organized as follows: Section~\ref{s:bimodal} is devoted to the construction of our multi-modal 
diffusion model. We investigate the properties of these model and contrast them to other 
existing bimodal diffusions, the double-well potential models in particular. In Section~\ref{s:inference} 
we discuss inference for the new class of bimodal diffusions emphasizing approximate likelihood inference
based on martingale estimating functions. Inference is further discussed when the bimodal diffusion 
is observed with measurement error. In Section~\ref{s:case} we apply a bimodal diffusion model to molecular
dynamics data in form of a reaction coordinate of the small Trp-zipper protein. Upon adjusting for measurement 
error we arrive at estimated folding rates that are realistic for this kind of protein. 
Section~\ref{s:concl} concludes.

\section{Multi-modal diffusions by transformation}
\label{s:bimodal}
%In this section we demonstrate how tractable multi-modal diffusions
%can be constructed using suitable transformations. We discus the basic properties
%of the transformed diffusion models and contrast them with other
%multi-modal diffusion models.

\subsection{Multi-modal distributions}
As a starting point we consider a bimodal density $f$. For instance,
$f$ could be the invariant density of the double-well potential
diffusion (\ref{e:dwpdif}) or the mixture of two unimodal densities
$f_1$ and $f_2$, 
\begin{equation}
\label{e:bidens}
f=\alpha\cdot f_1 + (1-\alpha)\cdot f_2, \quad 0<\alpha<1.
\end{equation}
This density will be the marginal invariant density of our bimodal diffusion.
As our main example, we consider the bimodal normal mixture density
$\alpha\cdot\phi(\cdot;\mu_1,\sigma_1^2)+(1-\alpha)\phi(\cdot;\mu_2,\sigma_2^2)$,
where the mean parameters $\mu_1$ and $\mu_2$ determine the location of the
two modes and the variance parameters $\sigma_1^2$ and $\sigma_2^2$ determine
the broadness of each mode. Some instances of this density are shown in 
Figure~\ref{f:bigauss} below. The bimodal normal density fits the protein reaction
coordinate data in our case study (Section \ref{s:case}) well. In other applications it 
might be more relevant to consider densities $f_1$ and $f_2$ that are constrained to a bounded interval
(this is the case for certain protein reaction coordinates), or that
are heavy-tailed or skew (e.g.\ financial data).

Note that if the mode points of the two unimodal densities are not located sufficiently 
far apart, then the mode points of the bimodal density are not identical to the mode points 
of the unimodal densities, or the mixture density might not be bimodal at all. 
This, however, is not a problem when applying the model to data where bimodality is manifest. 
Further, for particular choices of parameters, the bimodality of (\ref{e:bidens}) can easily 
be checked; its mode points can be found numerically. 

An important point of the paper is that multi-modal diffusions with more than two regimes 
can also be constructed by the method presented in this section by simply 
choosing for $f$ a multi-modal density. For instance, a tri-modal diffusion is
obtained for $f=\alpha_1f_1+\alpha_2f_2+\alpha_3f_3$, $\alpha_1, \alpha_2,
\alpha_3 \in (0,1)$, $\alpha_1+\alpha_2+\alpha_3=1$, where $f_i$,
$i=1,2,3$, are unimodal probability densities whose modes are
sufficiently separated. It is merely to simplify the presentation that
we describe only the bimodal model in detail.

\subsection{Bimodal diffusions}
In order to model a bimodal (multi-modal) diffusion, we initially consider a stationary
diffusion of general form,
\begin{equation}
\label{e:simpledif}
dX_t=\mu(X_t)dt + \sigma(X_t)dB_t,
\end{equation}
where $\{B_t\}$ is a Wiener process, and where we assume that the coefficients are sufficiently 
regular to ensure that a unique weak solution exists for any given initial condition. In principle
$\{X_t\}$ could be any diffusion, but we aim to construct bimodal diffusions for which 
statistical inference is relatively easy, so we are interested in cases where the diffusion $\{X_t\}$ 
is analytically tractable. This is for instance the case if $\{X_t\}$ is an ergodic Pearson diffusion 
as considered by \cite{fs:08}, see in addition \cite{ew:64} for an early account on these processes.

Recall that a stationary solution $\{X_t\}$ to the stochastic differential equation exists if
\begin{displaymath}
\int_{x^\#}^r s(x) dx = \int^{x^\#}_{\ell}s(x)dx = \infty
\quad \textnormal{ and }\quad
\int_{\ell}^r[s(x)\sigma^2(x)]^{-1}dx < \infty.
\end{displaymath}
where the state space is $(\ell,r)$ ($-\infty \leq \ell < r \leq \infty$), $x^\# \in (\ell,r)$ 
is arbitrary, and $s(x)$ is the density of the scale measure 
\begin{equation}
\label{scale}
s(x)=\exp \left( -2 \int_{x^\#}^x \frac{\mu(y)}{\sigma^2(y)} 
dy \right), \ \ \ x \in (\ell,r).
\end{equation}
Under these conditions the diffusion $\{X_t\}$ is ergodic with invariant probability density given by 
$\pi(x)=(\int_{\ell}^r\{s(y)\sigma^2(y)\}^{-1}dy)^{-1}\cdot\{s(x)\sigma^2(x)\}^{-1}$, see \cite{kt:81}.

Let $\Pi$ denote the cumulative distribution function of the invariant distribution of the diffusion 
$\{X_t\}$ given by (\ref{e:simpledif}), then a stationary diffusion with bimodal (multi-modal) 
invariant density can be obtained by transforming $\{X_t\}$ with the transformation
$$\tr=F^{-1}\circ \Pi$$ 
where $F^{-1}$ is the quantile function of the bimodal (multi-modal) density. 
The cumulative distribution function of the bimodal mixture density (\ref{e:bidens}) is obviously given by 
$F(y) = \alpha F_1(y) + (1-\alpha)F_2(y)$, where $F_1$ and $F_2$ are the cumulative distribution
functions of the two unimodal distributions, so for practical purposes $F^{-1}$ can easily be computed. 
The dynamics of the transformed diffusion $\{Y_t\}=\{\tr(X_t)\}$ are given by the stochastic differential 
equation,
\begin{eqnarray}
\label{e:trdif}
dY_t &=& \mu^{\tr}(Y_t)+\sigma^{\tr}(Y_t)dB_t,\\
\mu^{\tr}(y)&=&
\frac{2\mu(\tr^{-1}y)\pi(\tr^{-1}y)+\sigma^2(\tr^{-1}y)\pi'(\tr^{-1}y)}{2f(y)}
- \frac{\sigma^2(\tr^{-1}y)\pi^2(\tr^{-1}y)f'(y)}{2f(y)^3} \nonumber \\
\sigma^{\tr}(y)&=& \frac{\pi(\tr^{-1}y)\sigma(\tr^{-1}y)}{f(y)} \nonumber 
\end{eqnarray}
with $\tr^{-1}=\Pi^{-1}\circ F$ and where we have abbreviated $\tr^{-1}(y)$ as $\tr^{-1}y$.
\begin{figure}[hbtb]
\begin{center}
\includegraphics[height=5cm,width=5cm,angle=0]{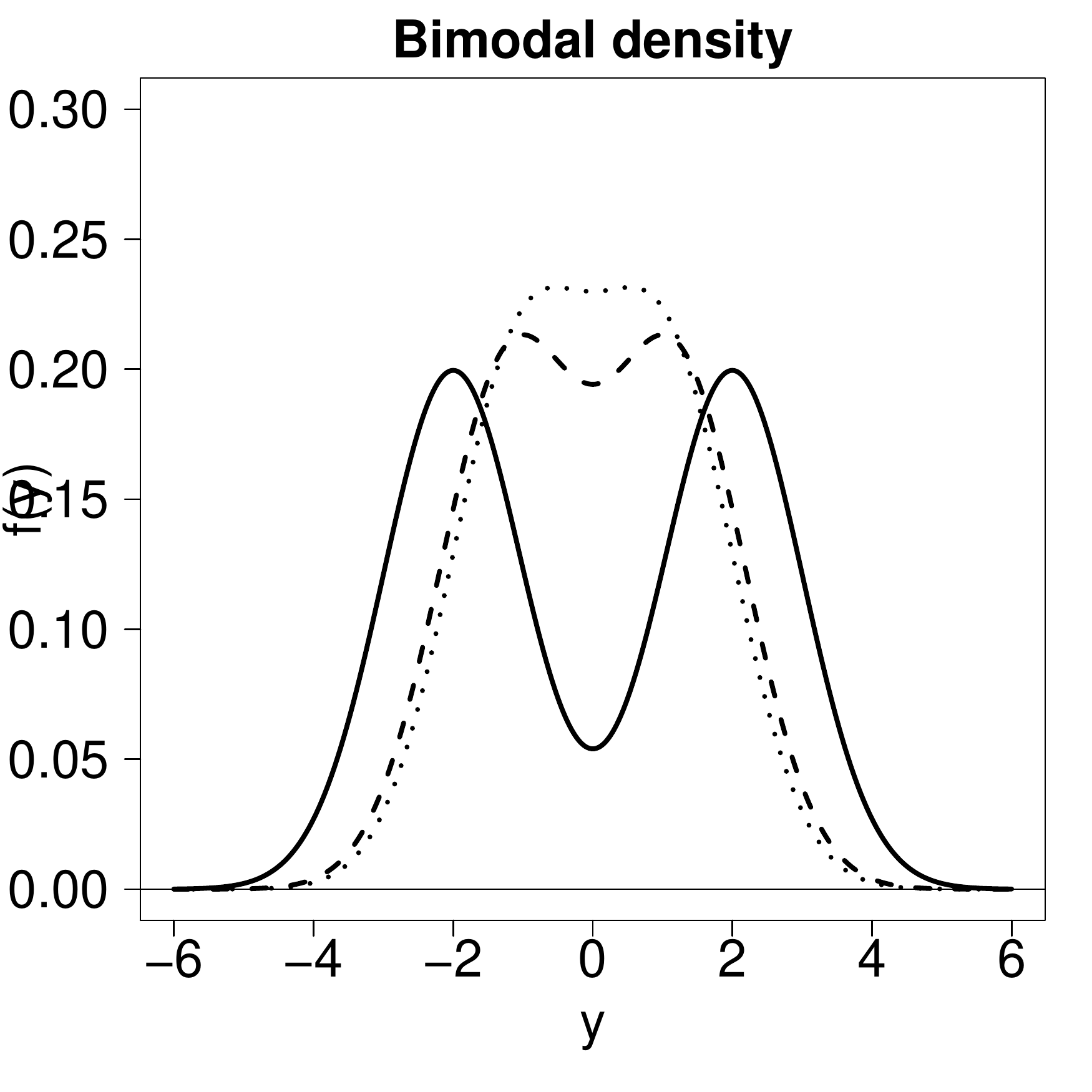}
\includegraphics[height=5cm,width=5cm,angle=0]{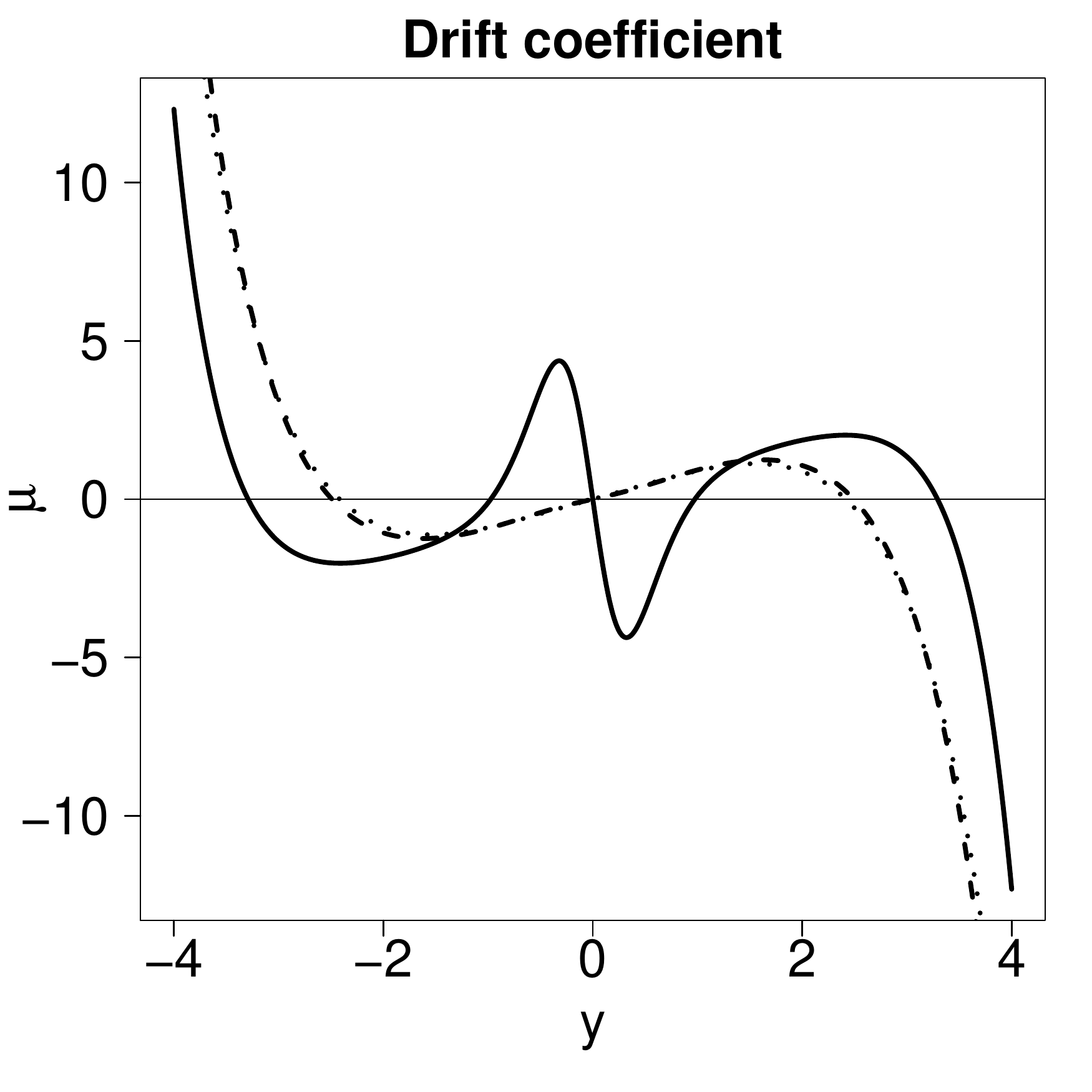}
\includegraphics[height=5cm,width=5cm,angle=0]{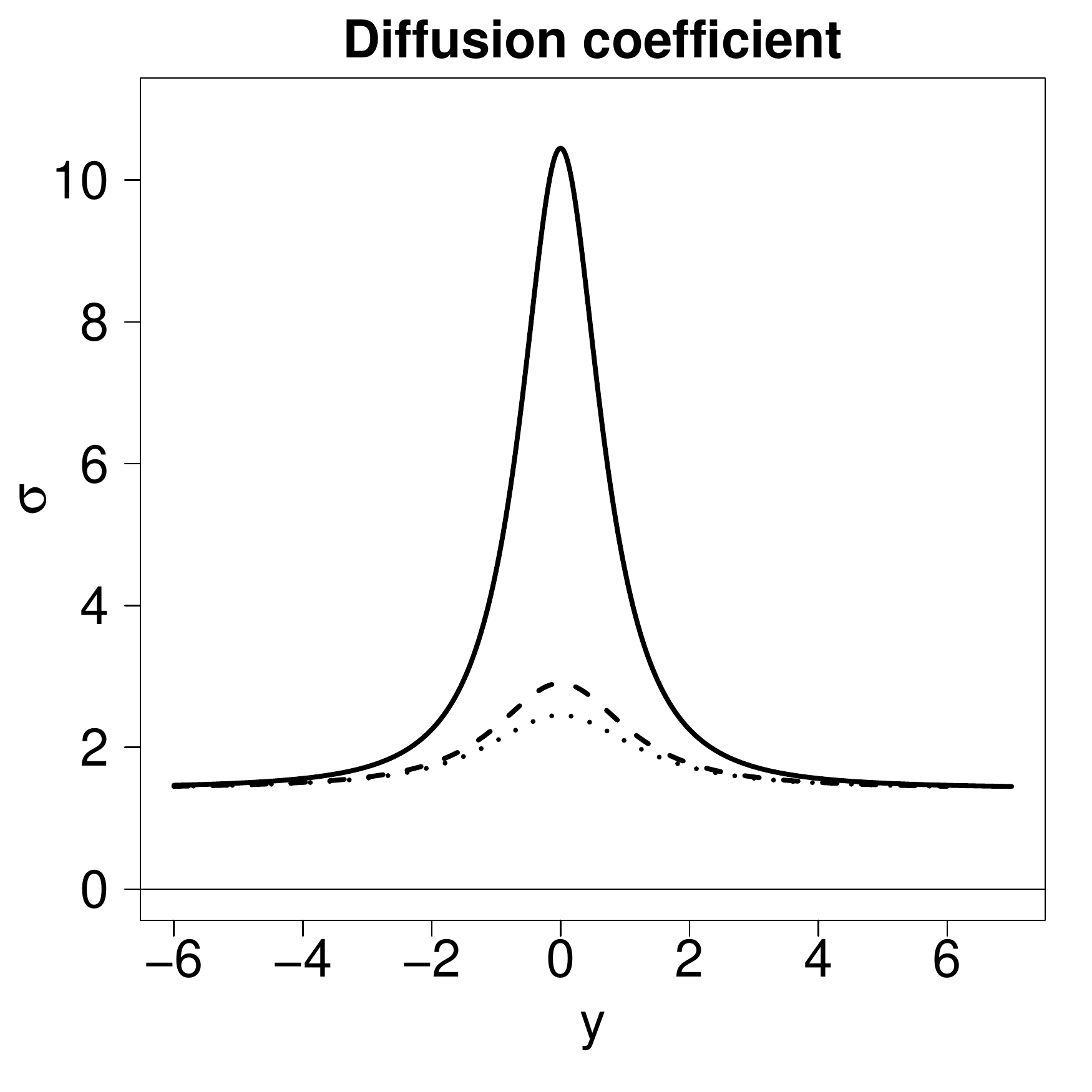}
\includegraphics[height=5cm,width=5cm,angle=0]{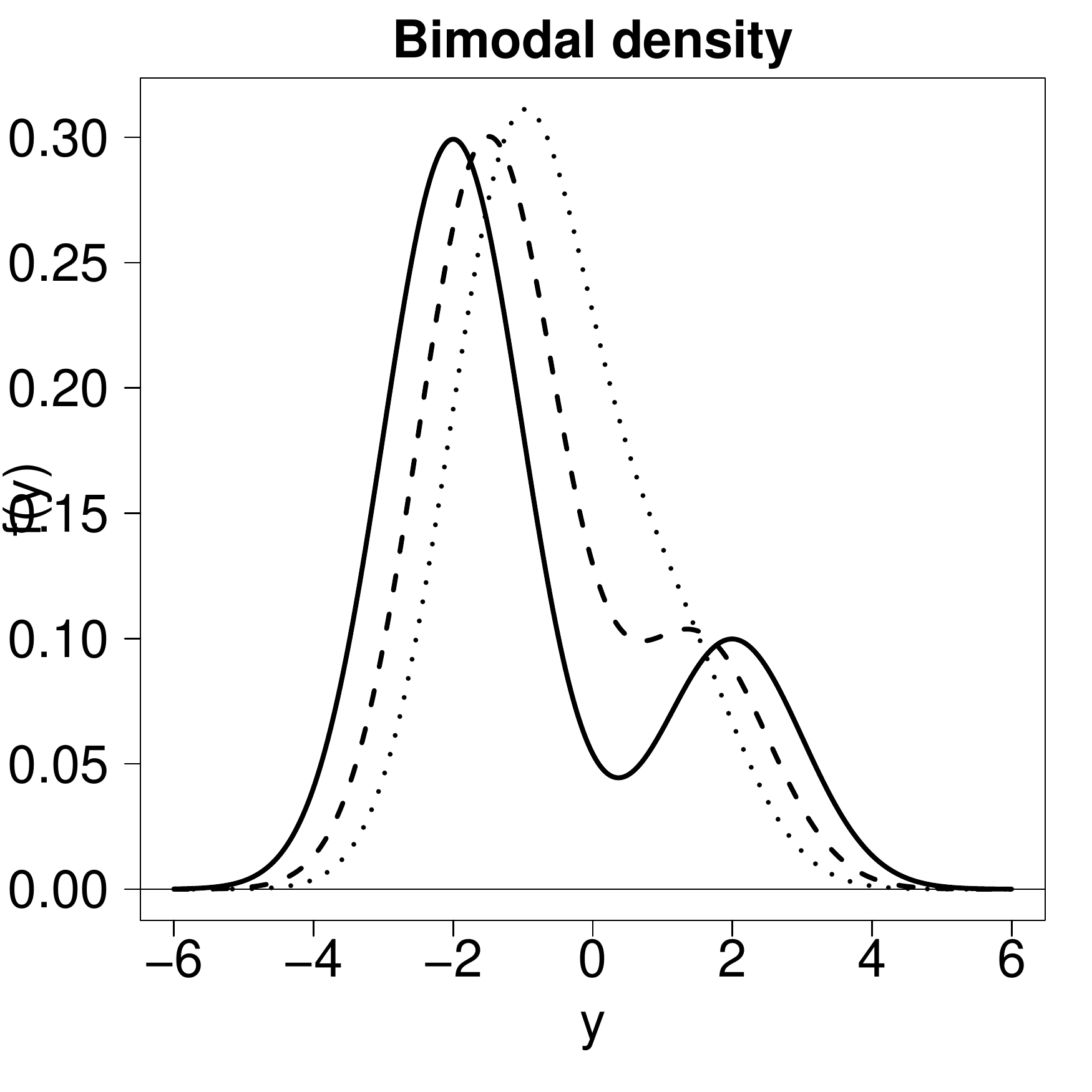}
\includegraphics[height=5cm,width=5cm,angle=0]{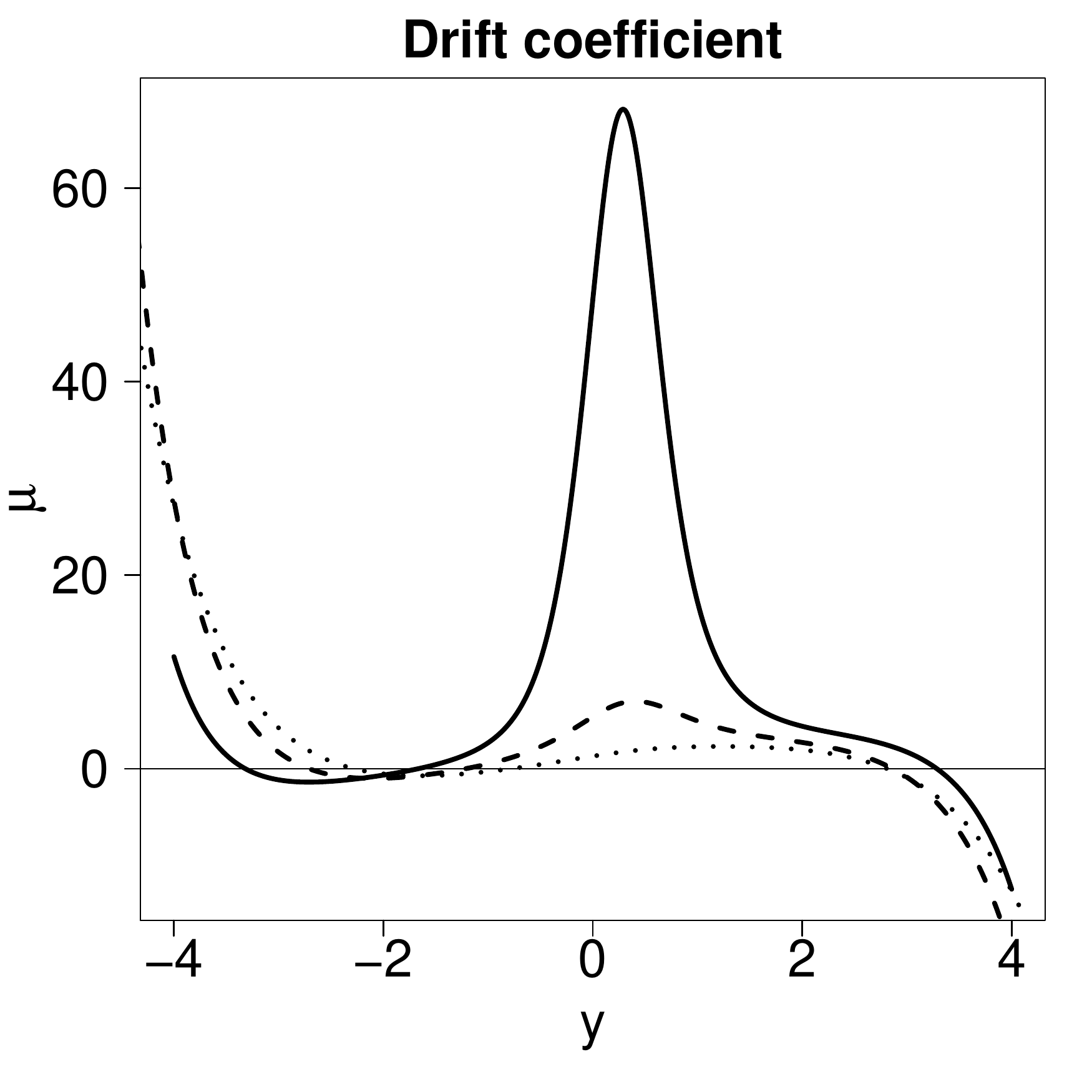}
\includegraphics[height=5cm,width=5cm,angle=0]{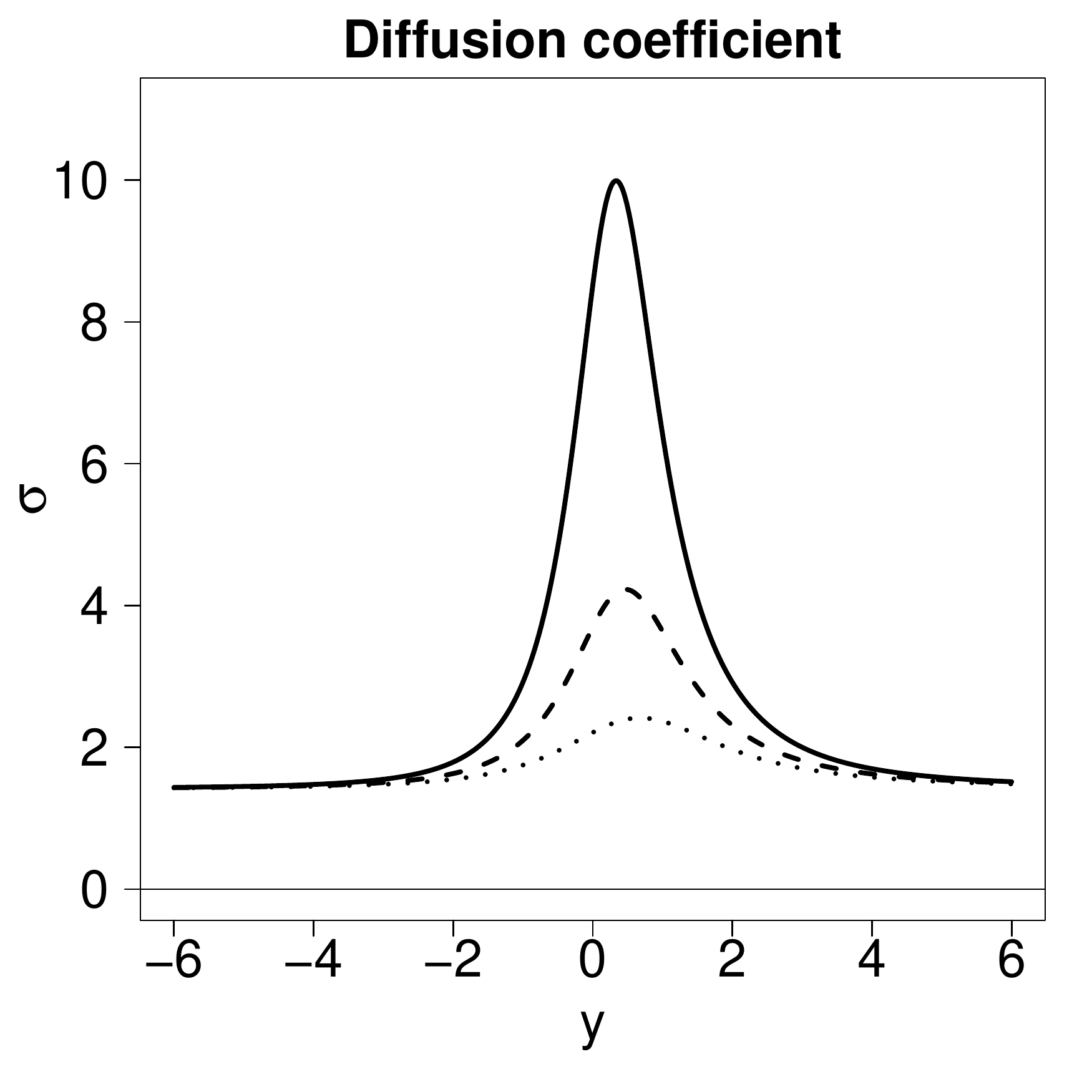}
\caption{Densities, drifts and diffusion coefficients of transformed Ornstein-Uhlenbeck processes 
(\ref{e:trou}) with parameters $\beta=1$ and invariant bimodal normal mixture densities given by 
$f(y)=\alpha\phi(y;\mu_1,\sigma_1)+(1-\alpha)\phi(y;\mu_1,\sigma_2)$
with $\sigma_1=\sigma_2=1$ and $-\mu_1=\mu_2=\mu$ varying as $\mu=1.05$ (solid lines), $\mu=1.5$ (dashed lines),
or $\mu=2$ (dotted lines). The upper panel shows symmetric densities ($\alpha=0.5$)
and the lower panel asymmetric densities ($\alpha=0.75$).}
\label{f:bigauss}
\end{center}
\end{figure}

\noindent
{\bf Example:} Transformation of the stationary Ornstein-Uhlenbeck process,
\begin{equation}
\label{e:oudif}
dX_t=-\nu X_tdt + \sqrt{2\nu}dB_t
\end{equation}
with invariant distribution equal to the standard normal distribution and autocorrelation function
$\rho(t)=e^{-\nu t}$, yields the diffusion
\begin{equation}
\label{e:trou}
dY_t = -2\nu\cdot\left(\frac{\tr^{-1}(Y_t)\varphi(\tr^{-1}Y_t)}{f(Y_t)}
+ \frac{\varphi^2(\tr^{-1}Y_t)f'(Y_t)}{2f(Y_t)^3}\right)dt 
+ \frac{\sqrt{2\nu}\varphi(\tr^{-1}Y_t)}{f(Y_t)}dB_t,
\end{equation}
where $\tr^{-1}=\Phi^{-1}\circ F$, and where $\varphi$ and $\Phi$ denote the density and the 
cumulative distribution function, respectively, of the standard normal distribution. Drifts and 
diffusion coefficients of some transformed Ornstein-Uhlenbeck process are shown in 
Figure~\ref{f:bigauss}. Note that the diffusion coefficients of the transformed Ornstein-Uhlenbeck 
processes peaks in-between the modes, which makes the model a good candidate for
protein reaction coordinates such as the one studied in Section \ref{s:case}. \hfill$\bullet$ 

Both the transformed Ornstein-Uhlenbeck process (\ref{e:trou}) and the general transformed 
diffusion (\ref{e:trdif}) inherits many properties from the simple diffusion $\{X_t\}$. 
The following results are straight-forward and thus stated without proof.

\begin{prop}
\label{t:props}
Let $f$ be the bimodal mixture density defined by (\ref{e:bidens}), and $\{X_t\}$ and
$\{Y_t\}$ the diffusions specified by (\ref{e:simpledif}) and (\ref{e:trdif}), 
respectively, then the following hold true.
\begin{enumerate}
\item If the densities $f_1$ and $f_2$ distributions have finite $k'th$ order moments
$\mu_k(f_1)$ and $\mu_k(f_2)$, then so has the bimodal mixture density;
$\mu_k(f)=\alpha\mu_k(f_1)+(1-\alpha)\mu_k(f_2)$. In particular, the mean and variance 
are $E(f)=\alpha E(f_1)+(1-\alpha) E(f_2)$ and 
\begin{displaymath}
\var(f)=\alpha\var(f_1)+(1-\alpha)\var(f_2)+\alpha(1-\alpha)\{E(f_1)-E(f_2)\}^2.
\end{displaymath}
\item If $\{X_t\}$ is stationary and ergodic, then so is $\{Y_t\}$.
\item The ($\alpha$, $\beta$ and $\rho$) mixing coefficients of $\{X_t\}$ and $\{Y_t\}$ satisfy 
$\kappa_t(X)=\kappa_t(Y)$, where $\kappa$ is either of $\alpha$, $\beta$, or $\rho$.
In particular if $\{X_t\}$ is $\kappa$-mixing (i.e.\ $\kappa_t\to 0$ as $t\to\infty$), 
then so is $\{Y_t\}$. 
%Recall that the mixing coefficients or a stationary Markov process are
%\begin{eqnarray*}
%\alpha_t(X)&=&\sup_{A,B\in\mathbf{B}}|P(X_0\in A, X_t\in B)-P(X_0\in A)P(X_t\in B)|,\\
%\beta_t(X)&=& E(\mathrm{ess}\sup_{A,B\in\mathbf{B}}|P(X_t\in B|X_0\in A)-P(X_t\in B)|),\\
%\rho_t(X)&=&\sup_{g,h\in L_2(\pi)}|\corr\{g(X_0),h(X_t)\}|.
%\end{eqnarray*}
We refer to \cite{dm:94} for further explanation. 
\item  If $g$ is an eigenfunction for the {\em infinitesimal generator} of the diffusion 
$\{X_t\}$ with eigenvalue $\lambda$, then $g\circ\tr^{-1}$ is an eigenfunction of
the generator of $\{Y_t\}$ with the same eigenvalue (the infinitesimal
generator of the diffusion (\ref{e:simpledif}) is the differential operator 
$\mathcal{L}=\mu\cdot\frac{d}{dx}+\frac{\sigma^2}{2}\frac{d^2}{dx^2}$, see \cite{kt:81}.
In particular it holds that $E\{g(\tr^{-1}Y_t)|Y_0\}=e^{-\lambda t}g(\tr^{-1}Y_0)$ under
mild regularity conditions, see \cite{ks:99}. 
\item If $\{p(x|x',\Delta)\}$ denote the transition densities of $\{X_t\}$  
(i.e.\ $p(x|x',\Delta)$ is the conditional density of $X_{t+\Delta}$ given $X_t=x'$), 
then the transition densities $\{q(y|y',\Delta)\}$ of $\{Y_t\}$ are given by 
$q(y|y',\Delta)=p(\tr^{-1}y|\tr^{-1}y',\Delta)\cdot f(y) / \pi(\tr^{-1}y)$.
\end{enumerate}
\end{prop}

One of the merits of the bimodal diffusion model is its capability of quantifying the shifts 
between its two regimes. These can be described by the passage times of the diffusion. 
We make use of this to estimate the folding and unfolding rate of the small Trp-zipper protein 
in Section \ref{s:case}.

The first passage time to $b$ of the general diffusion (\ref{e:simpledif}) is defined by 
$$T_X(b)=\inf\{t\geq 0\colon X_t=b\}$$
and given that $a<b$ the mean passage time from $a$ to $b$ is equal to
\begin{displaymath}
E(T_X(b)|X_0=a)=2\int_{\ell}^a\pi(x)dx\cdot\int_a^bs(x)dx + \int_a^b\pi(x)\int_x^bs(y)dydx
\end{displaymath}
where $s(x)$ is the scale density given by (\ref{scale}), see \cite{kt:81}.
Similarly, for the passage from $b$ to $a$ it holds that
\begin{displaymath}
E(T_X(a)|X_0=b)=2\int_b^r\pi(x)dx\cdot\int_a^bs(x)dx + \int_a^b\pi(x)\int_a^xs(y)dydx.
\end{displaymath}
The distributions of the passage times of the transformed diffusion are directly related to those 
of the underlying simple diffusion, as obviously $T_{Y}(b)=T_{X}(\tr^{-1}b)$. Hence we have the 
following result:

\begin{prop}
\label{t:passage}
Denote by $T_{X}$ a first passage time of the diffusion (\ref{e:simpledif})
and by $T_{Y}$ a first passage time of the transformed diffusion (\ref{e:trdif}),
then for any $a,b\in(\ell,r)$:
\begin{equation}
\label{e:tpassage}
( T_Y(b)|Y_0=a ) \stackrel{\mathcal{D}}{=} (T_X(\tr^{-1}b)|X_0=\tr^{-1}a),
\hspace{2mm}
(T_Y(a)|Y_0=b) \stackrel{\mathcal{D}}{=} (T_X(\tr^{-1}a)|X_0=\tr^{-1}b).
\end{equation}
%and for any $k$ such that $E(T_Y(b)^k)<\infty$ or $E(T_Y(a)^k)<\infty$ it holds that
%$$E(T_Y(b)^k|Y_0=a)=E(T_X(\tr^{-1}(b))^k|X_0=a), \hspace{1mm}
%E(T_Y(a)^k|Y_0=b)=E(T_X(\tr^{-1}(a))^k|X_0=b).$$
\end{prop}

\noindent
{\bf Example:} The mean passage times of the transformed
Ornstein-Uhlenbeck process, (\ref{e:trou}), between points $a<b$ are given by
\begin{equation}
\label{e:oupassage1}
E(T_Y(b)|Y_0=a) =\frac{\sqrt{2\pi}}{\nu}\int_{\tr^{-1}a}^{\tr^{-1}b}\Phi(x)e^{x^2/2}dx
\end{equation}
and
\begin{equation}
\label{e:oupassage2}
E(T_Y(a)|Y_0=b) =\frac{\sqrt{2\pi}}{\nu}\int_{\tr^{-1}a}^{\tr^{-1}b}\{1-\Phi(x)\}e^{x^2/2}dx.
\end{equation}
where $\Phi$ is the cumulative distribution function of the standard normal distribution.

In fact the passage times of the Ornstein-Uhlenbeck can be described in greater detail
as it is possible to find analytical expressions of their densities, see \cite{app:04}
and the references therein. Although involving series expansions and special functions, 
the densities can be computed by means of designated software. 
This could potentially be used to further describe the folding and unfolding of the
small Trp-zipper protein studied in Section~\ref{s:case}.\hfill$\bullet$

\subsection{Comparison to other bimodal diffusion models}
Double well potential models such as (\ref{e:dwpdif}) considered in the introduction
are the state-of-the-art models for protein reaction coordinates. It should be noted 
that the double-well potential diffusion model is quite flexible and can be made to match 
any particular bimodal density $f$ by specifying the potential as $V(x)=-\frac{\sigma^2}{2}\log f(x)$.
One might suspect that a diffusion of our type is a transformation of yet another diffusion with 
constant diffusion coefficient that moves in a double-well potential, and that this is 
the real underlying cause for the bimodality. However, we will demonstrate that this is not the case: 
Suppose a diffusion $\{Y_t\}$ with bimodal invariant density $f$ is constructed by 
transformation of a simple diffusion $\{X_t\}$ by the method in the previous section. 
To any diffusion of the general form (\ref{e:simpledif}) corresponds a unique transformation 
to a diffusion $\{\widetilde{X}\}$ with constant diffusion coefficient, the so-called Lamperti transformation, 
$\widetilde{X}_t=\int_{x^\#}^{X_t}\{\sigma(y)\}^{-1}dy$, see e.g.\ \cite{si:08}. Since
$\sigma^{\tr}(y)=\sigma\{\tr^{-1}y\}\cdot\tr'\{\tr^{-1}y\}$, it is not difficult to show 
that when applying the Lamperti transformation of the bimodal transformed diffusion 
$\{Y_t\}=\{\tau(X_t)\}$ to this process, then the resulting process is identical to 
$\{\widetilde{X}_t\}$, which is completely unrelated to the bimodal density $f$. 
Our approach to bimodal diffusions 
is thus genuinely different from the double-well potential models. In the new models the bimodality 
is not caused by a double-well potential; rather the bimodality is the product of an interplay 
between the smooth motion given by the drift and the random fluctuations given by the state-dependent 
diffusion coefficient.

\cite{yas:96} proposed a general nonlinear diffusion model specified by,
\begin{equation}
\label{e:nld}
dX_t=(\alpha_{-1}X_t^{-1}+\alpha_0+\alpha_1X_t+\alpha_2X_t^2)dt
+\sqrt{\beta_0+\beta_1X_t+\beta_2X_t^\gamma}dB_t,
\end{equation}
The parameter constraints under which (\ref{e:nld}) is stationary and ergodic are
complicated and we will not quote them here. The nonlinear diffusion model may display 
bimodality, but at the same time it contains simple unimodal models such as the Ornstein-Uhlenbeck
process. The nonlinear bimodal diffusion process produces bimodality in the same way as our new 
model does, although in the model (\ref{e:nld}) most of the action is in the drift. In comparison with the other 
classes of models, the nonlinear diffusion (\ref{e:nld}) has the disadvantage that in its general form it does not 
admit simple explicit estimating functions, like our transformed diffusions do, and that analytical likelihood 
approximations do not simplify for (\ref{e:nld}) as they do for the double-well potential models, see Section
\ref{s:inference} below. Also the nonlinear diffusion (\ref{e:nld}) is not quite as easy to simulate as the other models.

In regard to the protein folding problem, any of the three classes of models may be relevant
as different kinds of reaction coordinates display varying patterns of diffusion. For instance,
our new model seems apt at modeling reaction coordinates that measure fluctuations in Cartesian 
configuration space for which diffusion is increased in-between modes, e.g.\ as in \cite{bh:10}, 
whereas the nonlinear diffusion model is a better model for a reaction coordinates 
in which diffusion decrease monotonically towards the folded state, e.g.\ as in \cite{colw:07}.

We end this Section by remarking that bimodality can also be produced solely by 
effects of the random fluctuations determined by a state dependent diffusion 
coefficient, as is clear from the following example. Suppose $f$ is a bimodal 
density function for which the support is the real line. Then the diffusion
given by
\begin{equation}
\label{diffonly}
dX_t = \frac{\sigma}{\sqrt{f(X_t)}}dB_t
\end{equation}
is ergodic with invariant density $f$, see e.g.\ \cite{finstoc}. The
bimodality is produced by the fact that near the two mode-points the
random fluctuations are relatively small, so that the process tends to
stay near these points. When the process is far from the mode-points,
the random fluctuations are large and will quickly send the process to
other parts of the state space. Thus the same bimodal invariant distribution
can be attained for diffusions with very different drift and diffusion
coefficients and by completely different mechanisms. Our approach
combines the mechanisms of the double-well potential diffusions (motion
in an energy landscape), and of the pure diffusion models
(\ref{diffonly}) (state-dependent random fluctuations).

\section{Statistical inference:}
\label{s:inference}

\subsection{Discretely observed diffusion}
In what follows we discuss inference for a discretely observed
multi-modal diffusion of the type constructed in Section
\ref{s:bimodal}. We assume that observations are made at equidistant
time points $t_i=i\Delta$ for $i=0,\ldots,N$, where $\Delta^{-1}>0$ is 
the sampling frequency. We denote the observations by $y_0, \ldots,
y_N$. Further, we assume that the distribution of the underlying
simple diffusion $\{X_t\}$ is dependent on a $p_1$-dimensional vector
of parameters $\nu$ and that the bimodal density $f_\psi$ is dependent
on a $p_2$-dimensional vector of parameters $\psi$. \\ 
\\
{\bf Likelihood estimation}\\
Likelihood estimation is our preferred means of inference, but it is 
unfortunately rarely the case that the likelihood function of a diffusion
process is explicitly known. Therefore inference for diffusion models
is usually best carried out by means of approximate likelihood methods.
A noteworthy exception is the Ornstein-Uhlenbeck process and
transformations of it, for which the likelihood functions is explicitly
known. In particular this is true for our bimodal transformation of the
Ornstein-Uhlenbeck process for which the likelihood function is given by
\begin{displaymath}
L_n(\psi,\nu)=\prod_{i=0}^{n-1}\frac{1}{\sqrt{1-e^{-2\nu\Delta}}}\varphi\left(\frac{\tr^{-1}_{\psi,\nu}y_{i+1}
    -
    e^{-\nu\Delta}\tr^{-1}_{\psi,\nu}y_i}{\sqrt{1-e^{-2\nu\Delta}}}\right)\frac{f_{\psi}(y_{i+1})}{\varphi(\tr^{-1}_{\psi,\nu}y_{i+1})},
\end{displaymath}
where $\varphi$ denotes the standard normal density function.

Approximate likelihood inference is applicable to many diffusions by using 
\cite{yas:02}'s analytical approximation to the likelihood function. 
In a simulation study by \cite{jp:02} it was demonstrated that this particular 
likelihood approximation is superior to other approximate
likelihood methods both in terms of computing time  
and in terms of numerical accuracy. An implementation of this
method can be found in \cite{si:08}. The first step in the analytical likelihood approximation 
is to compute the Lamperti transform of the diffusion, see Section~\ref{s:bimodal}. Hence, 
with likelihood inference in mind it is desirable to choose an underlying
diffusion with a simple Lamperti transform. As demonstrated in Section~\ref{s:bimodal}
the process obtained when applying the Lamperti transform is invariant under transformation. 
Likelihood inference is therefore a natural choice for inference in the double-well potential
model (\ref{e:dwpdif}). Since the diffusion coefficient is already constant,
the first step in \cite{yas:02}'s approximation can be skipped.\\
\\
{\bf Martingale estimating functions}\\
Martingale estimating functions are another way of approximating 
likelihood inference. Specifically, they are unbiased approximations
to the score  function, see \cite{bjs:04}. For instance the quadratic
martingale estimating function, 
\cite{bs:95}, is a simple, yet often highly efficient, means for obtaining parameter 
estimates in a diffusion model, see \cite{ms:07}. If $H_N(\psi,\nu)$ denotes a general 
estimating function for the diffusion model (\ref{e:trdif}), then an estimate 
$(\hat{\psi},\hat{\nu})$ is obtained by solving the estimating equation $H_N(\psi,\nu)=0$.

For the Pearson diffusions we can not only find explicit expressions
for moments and conditional moments (to the extent these exist), but
also explicit polynomial eigenfunctions for the infinitesimal
generator. We can therefore find explicit eigenfunctions for
transformations of Pearson diffusions 
as shown in Proposition~\ref{t:props}. This in turn implies that we can find 
explicit martingale estimating functions of the type proposed by \cite{ks:99} 
for our bimodal diffusions. To be specific, suppose that $g_1(x;\nu), \ldots, g_k(x;\nu)$ 
are eigenfunctions of the infinitesimal generator of the underlying simple diffusion
with eigenvalues $\lambda_1(\nu), \cdots, \lambda_k(\nu)$. Then 
$g_1\circ\tr^{-1}, \ldots, g_k\circ\tr^{-1}$ are eigenfunctions of the
generator of $Y$ with the same eigenvalues. Therefore,
\begin{displaymath}
G_N (\nu,\psi) = \sum^N_{i=1} \sum_{j=1}^k w_j(y_i;\nu,\psi)
\left\{g_j(\tr_{\nu,\psi}^{-1}y_i;\nu) - e^{-\lambda_j(\nu)\Delta} 
g_j(\tr^{-1}_{\nu,\psi}y_{i-1};\nu)\right\}
\end{displaymath}
is a  martingale estimating function. In order to obtain an
approximation to likelihood inference, the ($p_1$+$p_2$)-dimensional
weight functions $w_j$ should be 
chosen optimally in the sense of \cite{gh:87}. If the eigenfunctions
are polynomials, then the optimal weight function can be found
explicitly as explained in \cite{fs:08} or \cite{MSSemstat}. This is
the case for the Pearson diffusions. If a Pearson diffusion is
transformed by a function $\tau$ that does not depend on the
parameters, then the optimal estimating function based on the
eigenfunctions of the transformed process is simply equal to the
estimating function obtained by inserting the data $\tau^{-1}y_i$ in
the optimal estimating function for the original Pearson diffusion,
see e.g.\ \cite{fs:08} or Theorem 1.19 in \cite{MSSemstat}. This fact
was used by  \cite{ls:07} to estimate the parameters in a model 
for exchange rates in a target zone.

For our bimodal diffusions, the situation is more complicated because
the transformation $\tau_{\nu,\psi}$ is parameter dependent. In the
following we extend the previous results to the case of a parameter
dependent transformation. To simplify the presentation, we consider the
case of polynomial eigenfunctions, $g_j(x;\nu)  = \sum_{\ell=0}^j
a^\nu_{j,\ell} \, x^\ell$. The optimal choice of the ($p_1$+$p_2)
\times k$-matrix of weights $w = (w_1, \ldots, w_k)$ is given by
\[
w^*(y;\nu,\psi) = B(y;\nu,\psi)V(y;\nu,\psi)^{-1},
\]
where the $k \times k$-matrix $V(y;\nu,\psi)$ is equal to $V_h$ in
Theorem 1.19 in \cite{MSSemstat} (with $\kappa(y) =
\tr_{\nu,\psi}^{-1}y$). The matrix $B$ is given by 
\[
B = \left( \begin{array}{c} B_h \\ 0_{p_2 \times k}\end{array} \right)
+ \tilde B,
\]
where the $p_1 \times k$-matrix $B_h$ is as in Theorem 1.19 in
\cite{MSSemstat} with only the $p_1$ derivatives w.r.t.\ $\nu$
included, $0_{p_2 \times k}$ is the $p_2 \times k$-matrix with all
entries equal to zero (this can be thought of as derivatives w.r.t.\
$\psi$), and the entries $\tilde B_{i,j}(y;\nu,\psi)$ are given by
$$\sum_{\ell = 1}^j a_{j,\ell}^\nu \, \ell \left[ E_{\nu,\psi} \left(
( \tr_{\nu,\psi}^{-1} Y_\Delta)^{\ell-1} \partial_{\theta_i}
\tr_{\nu,\psi}^{-1} Y_\Delta \, | \,  Y_0 = y
\right)  - e^{-\lambda_j (\nu)\Delta} \left( \tr_{\nu,\psi}^{-1} y 
\right)^{\ell-1}  \partial_{\theta_i} \tr_{\nu,\psi}^{-1} y \right],
$$
where $\theta = (\nu,\psi)$ ($i=1,\ldots,p_1+p_2, j=1,\ldots,k$).
Thus the optimal weights are explicit apart from the conditional
expectation in the expression for $\tilde B$, which can be determined
by simulation. Another solution is to expand the conditional
expectation and the exponential function in powers of $\Delta$, see
e.g.\ Lemma 1.10 in \cite{MSSemstat}. In this way the following
approximation to $\tilde B_{i,j}$ is obtained
\[
\tilde{\tilde{B}}_{i,j}(y;\nu,\psi) 
= \sum_{\ell = 1}^j a_{j,\ell}^\nu \, \ell \left[ L_{\nu,\psi} 
\left( (\tr_{\nu,\psi}^{-1})^{\ell-1} \partial_{\theta_i} 
\tr_{\nu,\psi}^{-1} \right) (y) - (\tr_{\nu,\psi}^{-1}y)^{\ell-1}
\partial_{\theta_i} \tr_{\nu,\psi}^{-1}y \right] \Delta, 
\]
where the differential operator $L_{\nu,\psi}$ is the infinitesimal 
generator of the diffusion $Y$ given by $L_{\nu,\psi} f(y) =
\mu^{\tr}(y) f'(y) + \frac12 \sigma^{\tr}(y) f''(y)$. The error made
when replacing $\tilde B$ by $\tilde{\tilde{B}}$ is of order $\Delta^2$,
so the loss of efficiency is small if the sampling frequency is
sufficiently high. \\

\noindent
{\bf Example:} The Ornstein-Uhlenbeck process (\ref{e:oudif}) has
(among many others) the
eigenfunctions $x$ and $x^2-1$, which for the untransformed process
gives a quadratic martingale estimating function. The corresponding
eigenvalues are $\nu$ and $2 \nu$. In this case
$\tr^{-1}_\psi=\Phi^{-1}\circ F_\psi$, where $\Phi^{-1}$ is the
quantile function of the standard normal distribution. Thus
$\tr_\psi^{-1}$ does not depend on the parameter $\nu$ of the
underlying Ornstein-Uhlenbeck process. This implies some
simplifications, e.g.
\[
\partial_\psi \tr^{-1}_\psi y = \frac{\partial_\psi F_\psi(y)}
{\varphi(\tr^{-1}_\psi y)}.
\]
and $\tilde B_{1j}=0$. For $j=1$, the other entries of $\tilde B$
are
\[
E_{\nu,\psi}\left( \partial_\psi \tr^{-1}_\psi Y_\Delta \, | \,
  \tr^{-1}_\psi Y_0 = y \right) - e^{-\nu\Delta} \partial_\psi
\tr^{-1}_\psi y.
\]

When $f(\cdot;\psi)$ is a bimodal mixture of two normal densities, 
$\psi=(\alpha,\mu_1,\sigma_1,\mu_2,\sigma_2)$ and the derivatives
$\partial_\psi F_\psi$ are easily found. They take the form
\begin{displaymath}
\partial_{\psi}F_\psi (y)=\left(\begin{array}{c}
\Phi(y;\mu_1,\sigma_1)-\Phi(y;\mu_2,\sigma_2)\\
 \rule{0mm}{4mm} -\frac{\alpha}{\sigma_1}\cdot\varphi(y;\mu_1,\sigma_1)\\
 \rule{0mm}{6mm} -\frac{\alpha(y-\mu_1)}{\sigma_1^2}\cdot
\varphi(y;\mu_1,\sigma_1)\\
 \rule{0mm}{4mm} -\frac{1-\alpha}{\sigma_2}\cdot\varphi(y;\mu_2,\sigma_2)\\
 \rule{0mm}{6mm} -\frac{(1-\alpha)(y-\mu_2)}{\sigma_2^2}\cdot
\varphi(y;\mu_2,\sigma_2)
\end{array}\right).
\end{displaymath}

\hfill$\bullet$\\
\\
{\bf Asymptotics}\\
Under weak regularity conditions, the maximum likelihood estimator and
the estimator obtained from the martingale estimating function $G_N$ are
consistent and asymptotically normal as $N \rightarrow \infty$. 
This follows from standard
asymptotic results, e.g.\ Theorem 1.2 in \cite{MSSemstat}. The
asymptotic variance of $\sqrt{N}(\hat \theta_N - \theta_0)$ can be
estimated by the inverse observed information in the case of the
maximum likelihood estimator and by the inverse of
\[
\hat J_N = \frac1N \sum_{i=1}^N B(y_i;\hat \nu_N, \hat \psi_N)
V(y_i;\hat \nu_N,\hat \psi_N)^{-1} B(y_i;\hat \nu_N, \hat \psi_N)^T,
\]
where $^T$ denotes transposition (the observed Godambe information),
in the case of $G_N$ with the optimal weights $w^*$. If the underlying
process is a Pearson diffusion, the regularity conditions on the 
parts of $G_N$ that are given by the corresponding optimal martingale
estimating function for the Pearson diffusion
$\{\tr_{\nu,\psi}^{-1}(Y_t)\}$ can be verified as in \cite{fs:08}
provided that the Pearson diffusion is ergodic with finite moments of
order $2k$. To treat the contribution to the estimating function from
$\tilde B$ (or $\tilde{\tilde{B}}$), additional conditions are needed
on the smoothness of the functions $\partial_{\theta_i}
\tr_{\nu,\psi}^{-1}$ and on the moments of $\tr_{\nu,\psi}^{-1} Y_t$
and $\partial_{\theta_i} \tr_{\nu,\psi}^{-1} Y_t$.

\subsection{Diffusion observed with measurement error}
In biological applications it is often not possible to measure the phenomenon of interest
without additional measurement error. Therefore we further discuss inference when a bimodal 
(multi-modal) diffusion is observed with error. For instance, the protein reaction coordinate 
considered in Section \ref{s:case} is much better fitted by a diffusion-plus-error model than by 
a plain diffusion model (even though the error in the protein reaction coordinate is not measurement 
error in its strict sense, but rather reflects local features of the folding funnel of the protein).

Suppose that instead of observing $\{Y_t\}$, we observe $\mathrm{z}(Y_{t_i},\varepsilon_{t_i})$, 
where $\mathrm{z}$ is a known function, and where $(\varepsilon_t)$ is a stationary, normally 
distributed and possibly correlated error process. For simplicity we consider the additive error model
\begin{equation}
\label{e:sumdif}
Z_t = Y_t + \varepsilon_t, \quad d\varepsilon_t = -\kappa\varepsilon_tdt + \sqrt{2\kappa\gamma^2}dW_t,
\end{equation}
where  $(\varepsilon_t)$ is an Ornstein-Uhlenbeck process with marginal $\mathcal{N}(0,\gamma^2)$-distribution 
and exponentially decaying autocorrelation function $\rho_{\varepsilon}(t)=\exp(-\kappa t)$. 
Note that the methods considered in this Section can easily be extended to other error processes 
/ other correlation functions.\\

\noindent
{\bf Example:} We consider again the bimodal Ornstein-Uhlenbeck model (\ref{e:trou}). The additional error complicates 
the statistical analysis as the observed process is no longer a Markov process.
Nevertheless the model (\ref{e:sumdif})  is still a tractable model in many regards as 
we will now demonstrate. The stationary marginal density of $\{Z_t\}$ is the bimodal normal mixture density
\begin{displaymath}
f(x) = \alpha\cdot\phi(x;\mu_1,\sigma_1^2+\gamma^2)
+(1-\alpha)\cdot\phi(x;\mu_2,\sigma_2^2+\gamma^2).
\end{displaymath}
Hence, consistent (though not efficient) estimates of $\alpha$,
$\mu_1$, $\mu_2$, $\sigma_1^2+\gamma^2$, and $\sigma_2^2+\gamma^2$ can
be obtained by maximizing the marginal likelihood function, i.e.\ by
pretending that observations are i.i.d.. Upon fixing the above
parameters at their estimates, the remaining parameters can be
estimated by a least squares fit of the theoretical to the
empirical autocorrelation function. The autocorrelation function of
$\{Z_t\}$ is given by 
\begin{displaymath}
\corr(Z_s,Z_{s+t}):=\rho_Z(t)=(1-\beta)\rho_Y(t)+\beta\exp(-\kappa t),
\end{displaymath}
where $\beta$ is the proportion of error variance in the marginal distribution, i.e.\
\begin{displaymath}
\beta=\frac{\var(\varepsilon_t)}{\var(Z_t)}=\frac{\gamma^2}
{\alpha(\sigma_1^2+\gamma^2)+(1-\alpha)(\sigma_2^2+\gamma^2)+\alpha(1-\alpha)(\mu_1-\mu_2)^2}
\end{displaymath}
and where $\rho_Y$ is the autocorrelation function of $\{Y_t\}$. The autocorrelations $\rho_Y(t_i)$ are 
not explicitly known, but due to the tractability of the bimodal diffusion they can easily be simulated. 
Initial values for the fit can be found by using that presumably
$\nu << \kappa$,  
so that $\rho_Z(t)\approx (1-\beta)+\beta\exp(-\kappa t)$ as $t\to 0$ and 
$\rho_Z(t)\approx(1-\beta)\rho_Y(t)$ as $t\to\infty$.\hfill$\bullet$\\

\noindent
More efficient estimators for the diffusion with error-model can be obtained from 
approximate likelihood methods, which however are all computationally intensive.
To name a few, we refer to \cite{cps:06} for Markov chain Monte Carlo methods and 
to \cite{fernando} for estimation based on the EM-algorithm.  
Inference can also be based on the prediction-based estimating functions, see \cite{ms:00} 
and \cite{ms2011}, although the joint moments in this context would have to be simulated.
The tractability of the latent underlying Ornstein-Uhlenbeck processes simplifies
all of the above mentioned methods (since it can be simulated exactly),
but only to a certain extent. Efficient estimation is the topic of ongoing research
which we will not pursue any further in the present article.

\section{Case study: The small Trp-zipper protein}
\label{s:case}
As an example, we consider molecular dynamics data in form of the L-reaction coordinate of 
the small Trp-zipper protein. The (high-dimensional) dynamics of the protein was simulated 
from the monte carlo algorithm \cite{sb:12} using the PHAISTOS software package \cite{wb:13}. 
Subsequently the L-reaction coordinate which measures the total distance to a folded reference was computed
resulting in a univariate time series. For the analysis we consider a subsample of 
20,000 observations corresponding to a sampling frequency of $\Delta^{-1}=$1/nsec.

\begin{figure}[httb]
\begin{center}
\includegraphics[height=5cm,width=7cm]{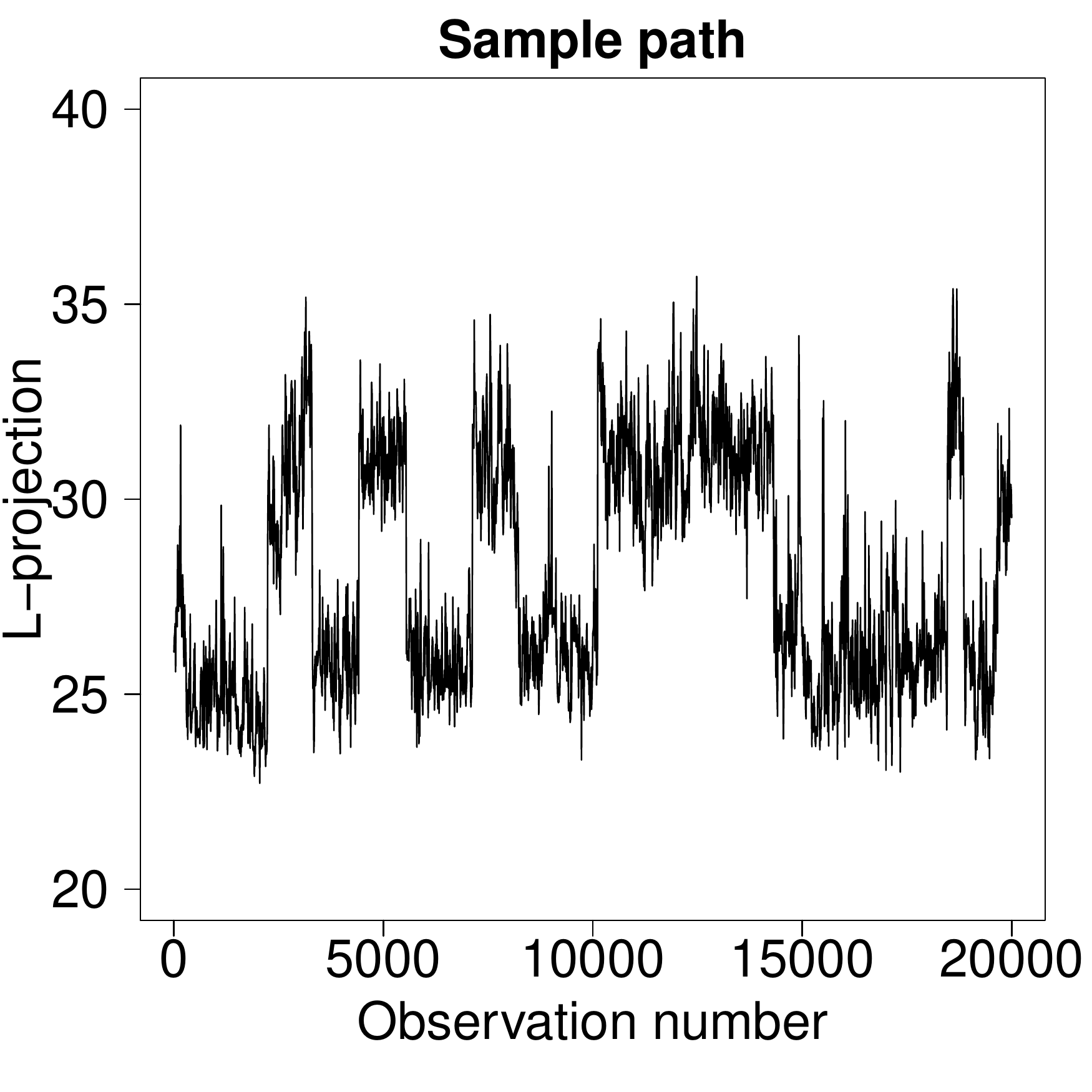} 
\includegraphics[height=5cm,width=7cm]{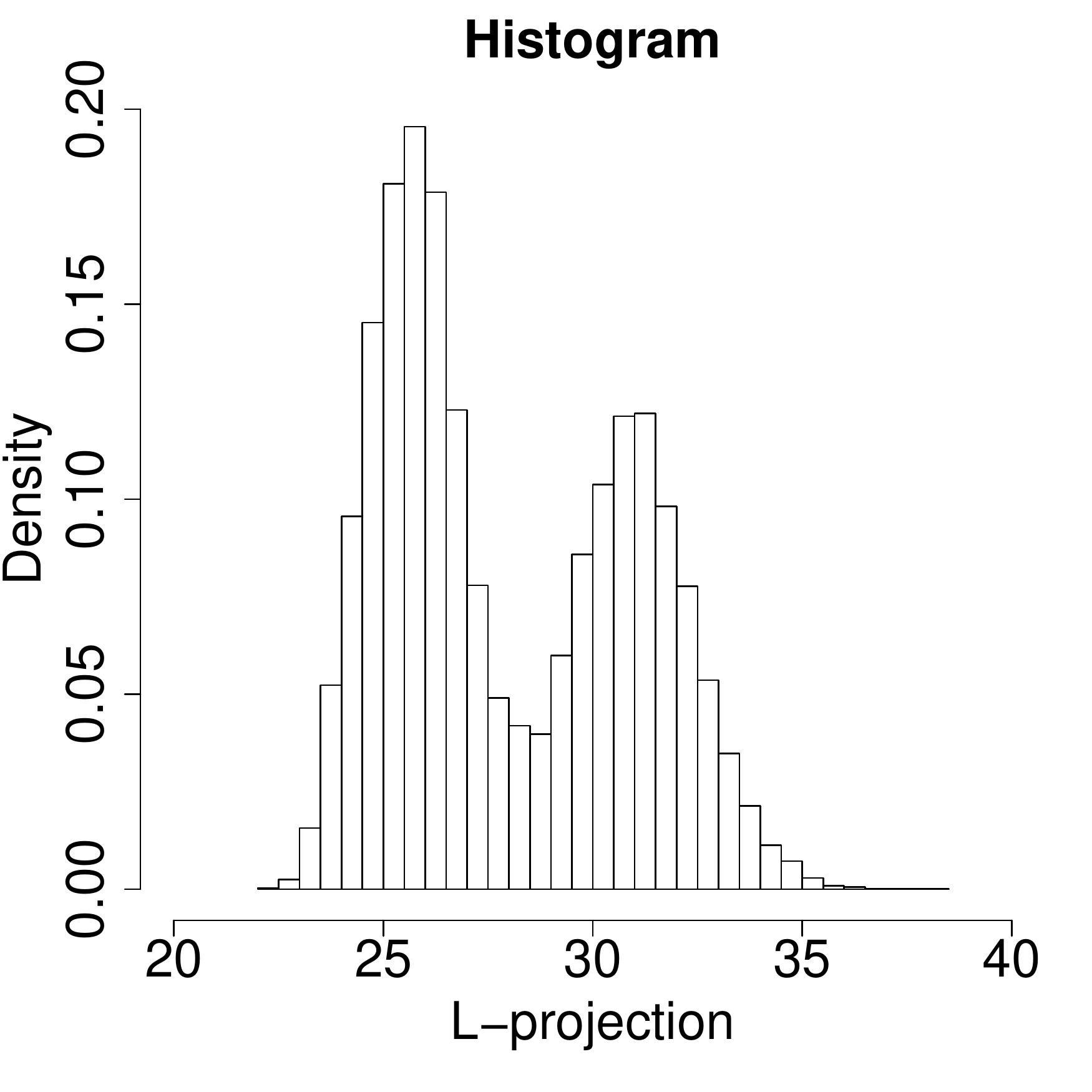} 
\includegraphics[height=5cm,width=7cm]{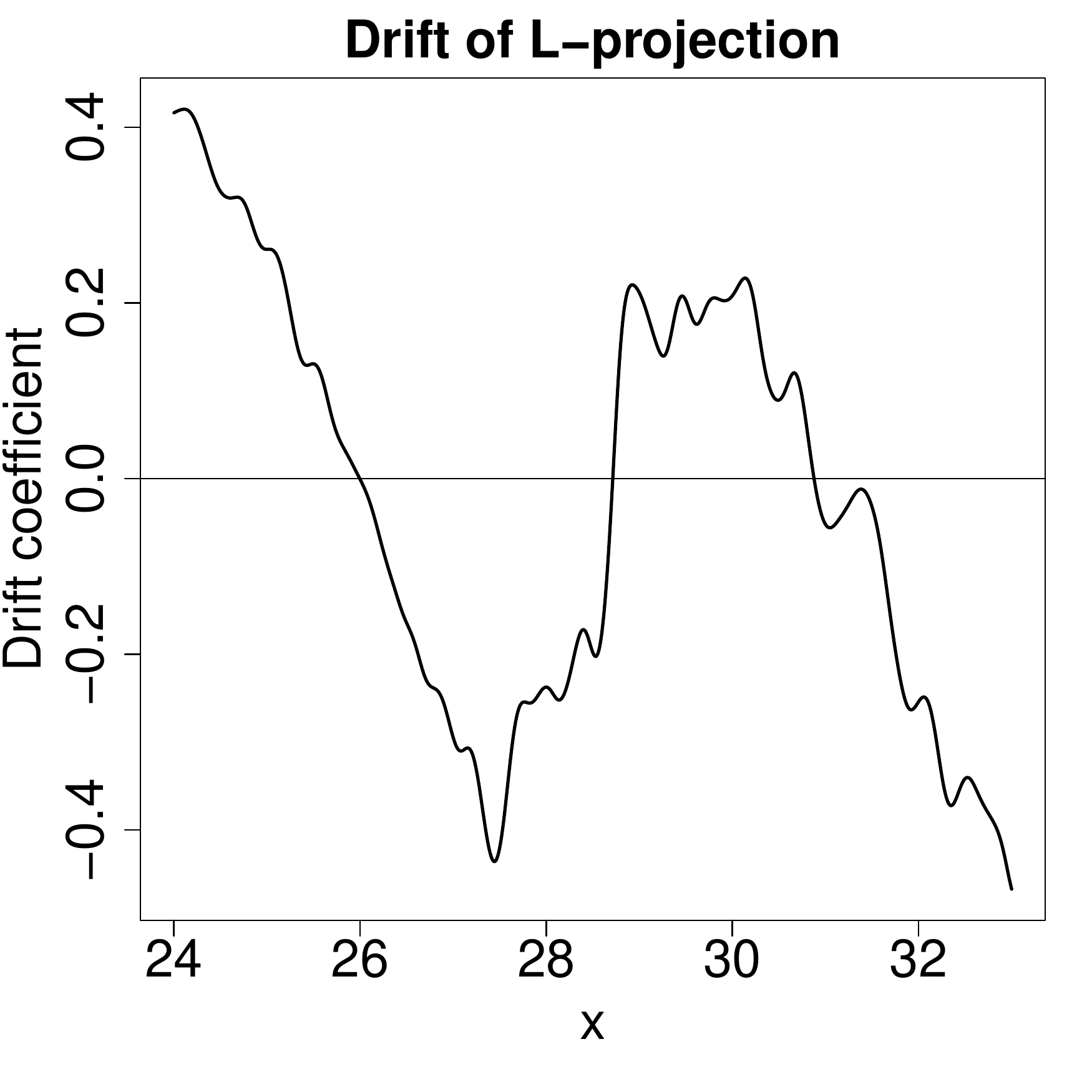} 
\includegraphics[height=5cm,width=7cm]{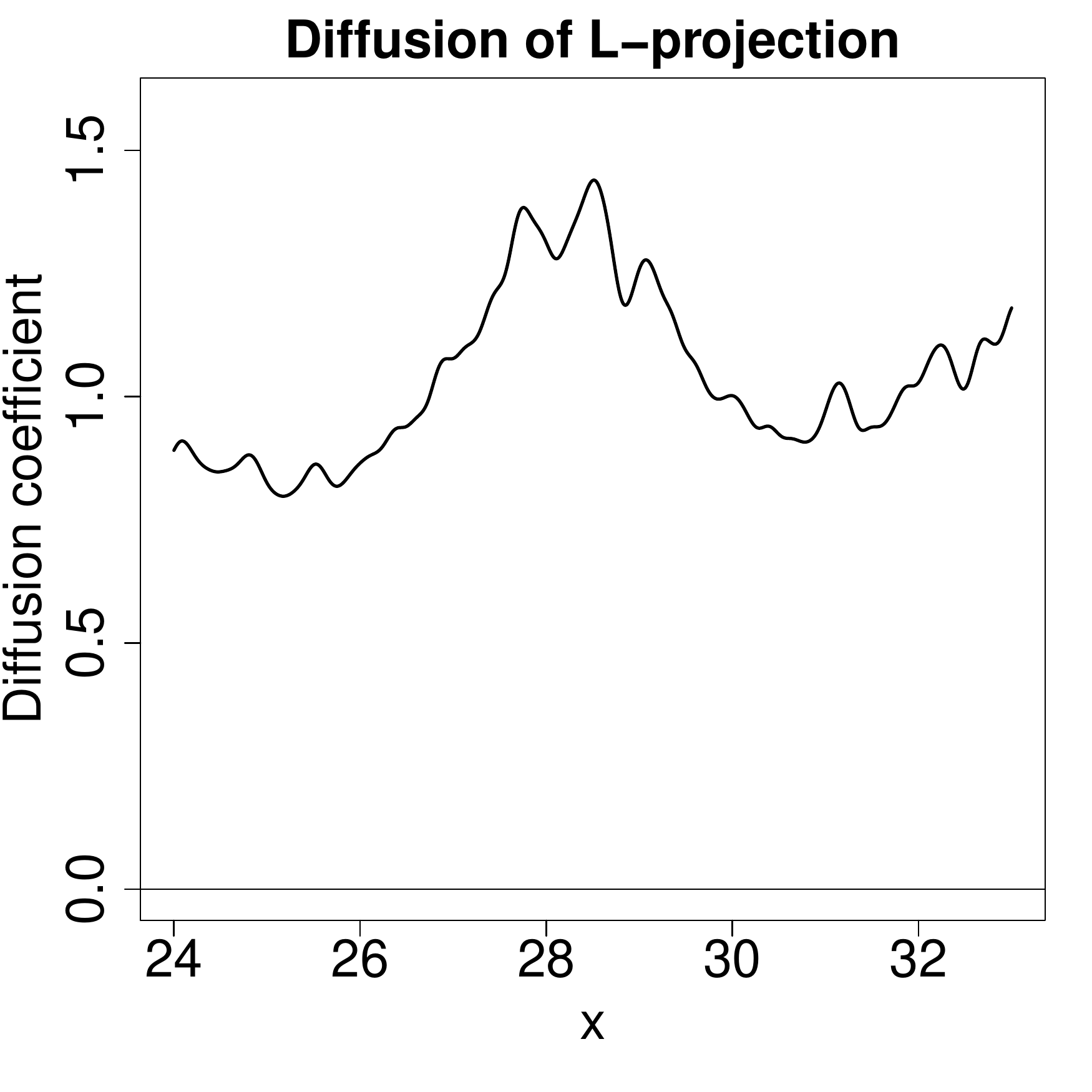} 
\caption{Sample path, sample histogram, and nonparametric estimates of drift and diffusion 
coefficient for 20.000 observations of the $L$-reaction coordinate of the small Trp-zipper protein.
The nonparametric estimators are local linear estimators, see \cite{fz:03}, 
evaluated at a bandwidth equal to 0.1.}
\label{f:Lprotein}
\end{center}
\end{figure}

The sample path of the $L$ reaction coordinate (Figure~\ref{f:Lprotein}) 
clearly reflects the two conformal states of the small Trp-zipper protein, 
{\em folded} and {\em unfolded}. The main interest is to estimate the folding and unfolding 
rates of the protein, and this can be achieved by application of a bimodal diffusion model where 
the rates of switching between the folded and unfolded state correspond to the mean passage times 
between the modes of the invariant bimodal density. 

\begin{figure}[httb]
\begin{center}
\includegraphics[height=5cm,width=7cm]{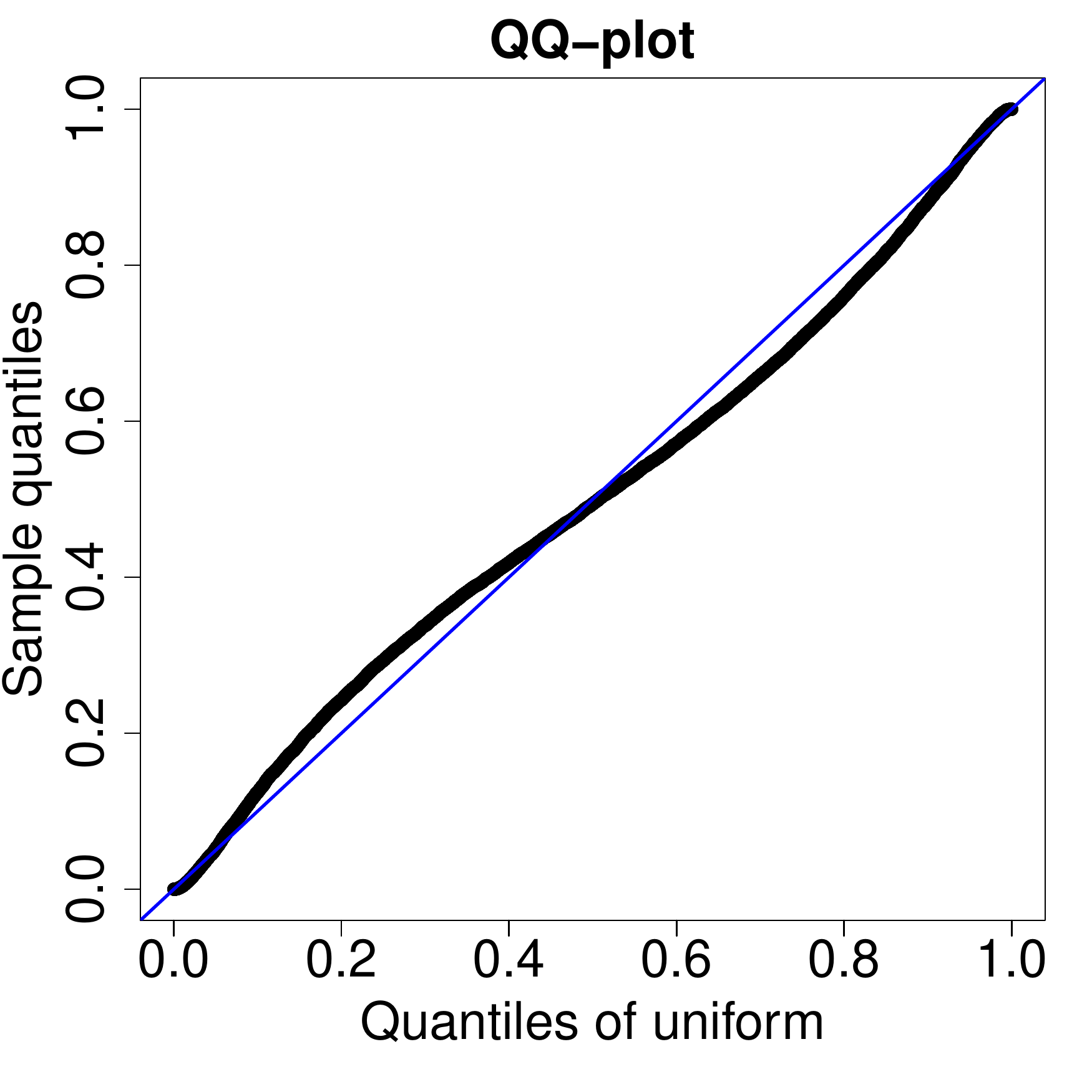} 
\includegraphics[height=5cm,width=7cm]{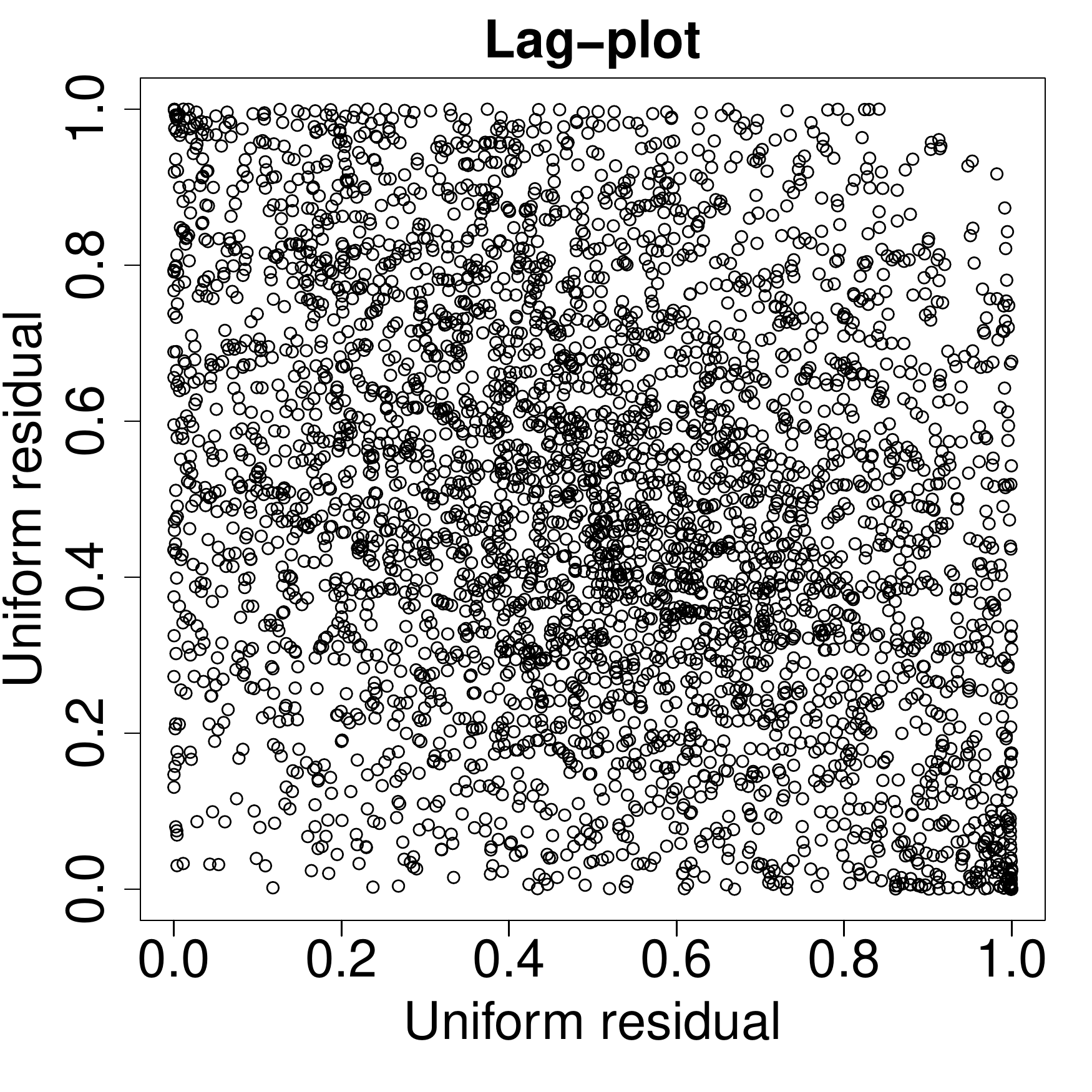} 
\caption{Diagnostics for the fit of the bimodal transformation of the
Ornstein-Uhlenbeck process to 20.000 observations of the $L$-projection
of the small Trp-zipper protein. The left panel shows the QQ-plot
of the uniform residuals and the right panel shows a lagplot of
consecutive uniform residuals.}
\label{f:diagLprotein}
\end{center}
\end{figure}

The histogram (Figure~\ref{f:Lprotein}) is well fitted by a bimodal normal mixture density. Further, 
the nonparametric estimates of drift and diffusion coefficient appear similar in shape to those of the 
bimodal transformation of the Ornstein-Uhlenbeck process (\ref{e:trou}). We note in particular, that the 
data display increased volatility in-between the modes. Similar patterns have been observed in other 
protein reaction coordinates, see e.g.\ \cite{bh:10}. All in all, the bimodal transformation 
of the Ornstein-Uhlenbeck process is a good candidate model for the data. We fit the model 
making use of the explicit likelihood function which yields the following parameter 
estimates: 
\begin{displaymath}
\hat{\alpha}=0.27,\quad
\hat{\mu}_1=25.41,\quad
\hat{\mu}_2=29.02,\quad
\hat{\sigma}_1=1.36,\quad
\hat{\sigma}_2=2.59,\quad\textnormal{and}\quad
\hat{\nu}=0.071.
\end{displaymath}
The lower and upper mode of the system are estimated by $\hat{l}=\hat{\mu}_1$ and $\hat{u}=\hat{\mu}_2$. 
The passage times of the model are given by (\ref{e:oupassage1}) and (\ref{e:oupassage2}). Thus we
find the estimated folding and unfolding rates of the protein:
\begin{eqnarray*}
\widehat{E}(T_{\ell}^Y|Y_0=u) &=& \frac{\sqrt{2\pi}}{\hat{\nu}}
\int_{\widehat{\tr^{-1}\mu_1}}^{\widehat{\tr^{-1}\mu_2}}\{1-\Phi(x)\}e^{x^2/2}dx
=28.8,\\ 
\widehat{E}(T_{u}^Y|Y_0=\ell) &=& \frac{\sqrt{2\pi}}{\hat{\nu}}
\int_{\widehat{\tr^{-1}\mu_1}}^{\widehat{\tr^{-1}\mu_2}}\Phi(x)e^{x^2/2}dx =18.5.
\end{eqnarray*}
Thus folding and unfolding should occur on average once in about
fifteen to thirty nsec. These estimates of the folding and unfolding
rates are unfortunately not realistic. We inspect the uniform
residuals (Figure~\ref{f:diagLprotein}) to check whether the  fit of
the bimodal transform of the Ornstein-Uhlenbeck process is
satisfactory. 

The fit to the uniform distribution is reasonable, but not perfect,
and the residuals appear to be negatively correlated, which might
indicate a misfit of the model. To check whether it is at all
plausible that the data was generated by a Markov process, we further
applied the nonparametric test of the Markov hypothesis of
\cite{asfj:10}. From this we concluded that it is not likely that the
protein data was generated by a Markov  
process ($P<0.0001$). Note that since the convergence of the Markov test statistic to its asymptotic 
chi-square distribution is poor, the p-value was computed by fully non-parametric bootstrapping
as described in \cite{jlf:12}. 

\begin{figure}[httb]
\begin{center}
\includegraphics[height=5cm,width=7cm]{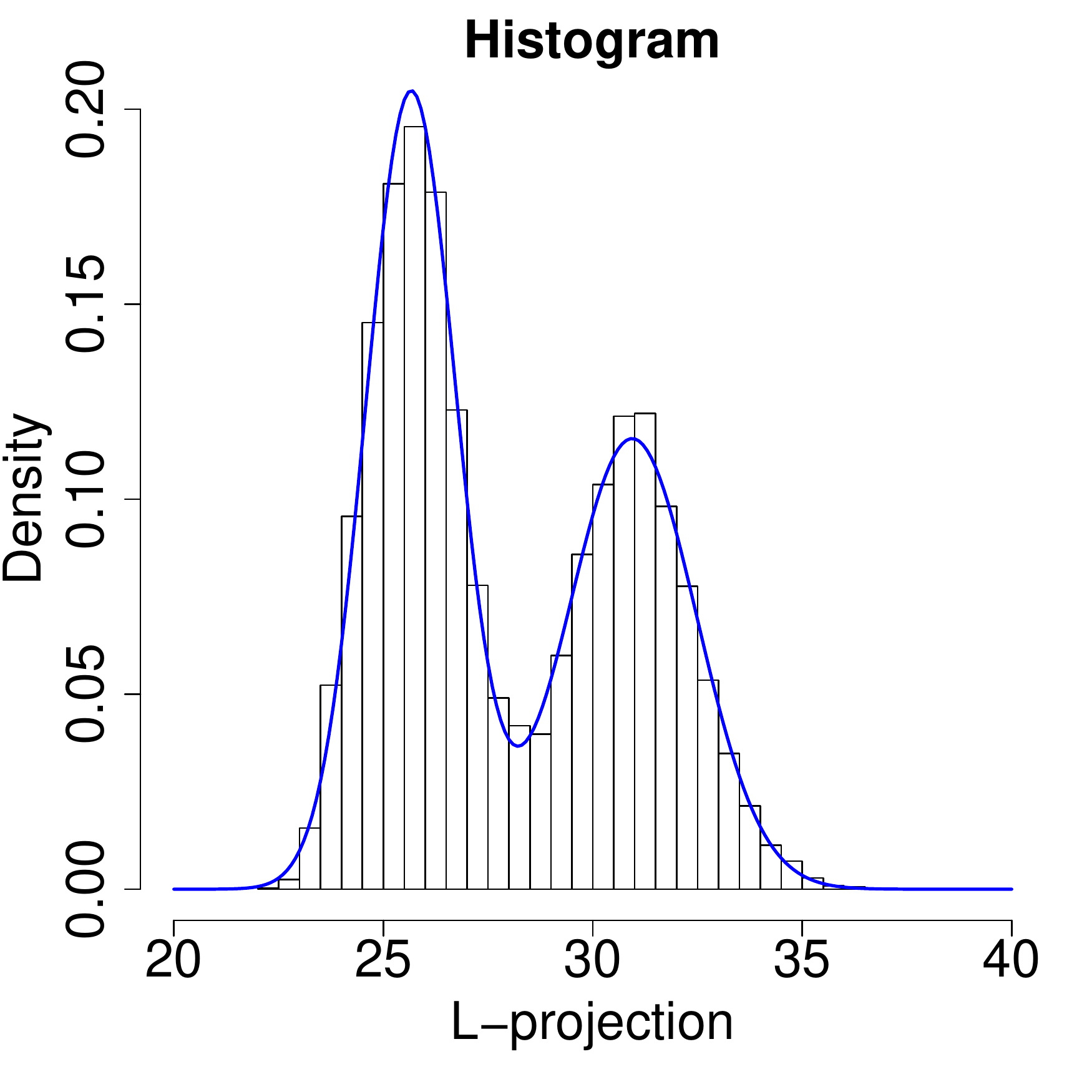} 
\includegraphics[height=5cm,width=7cm]{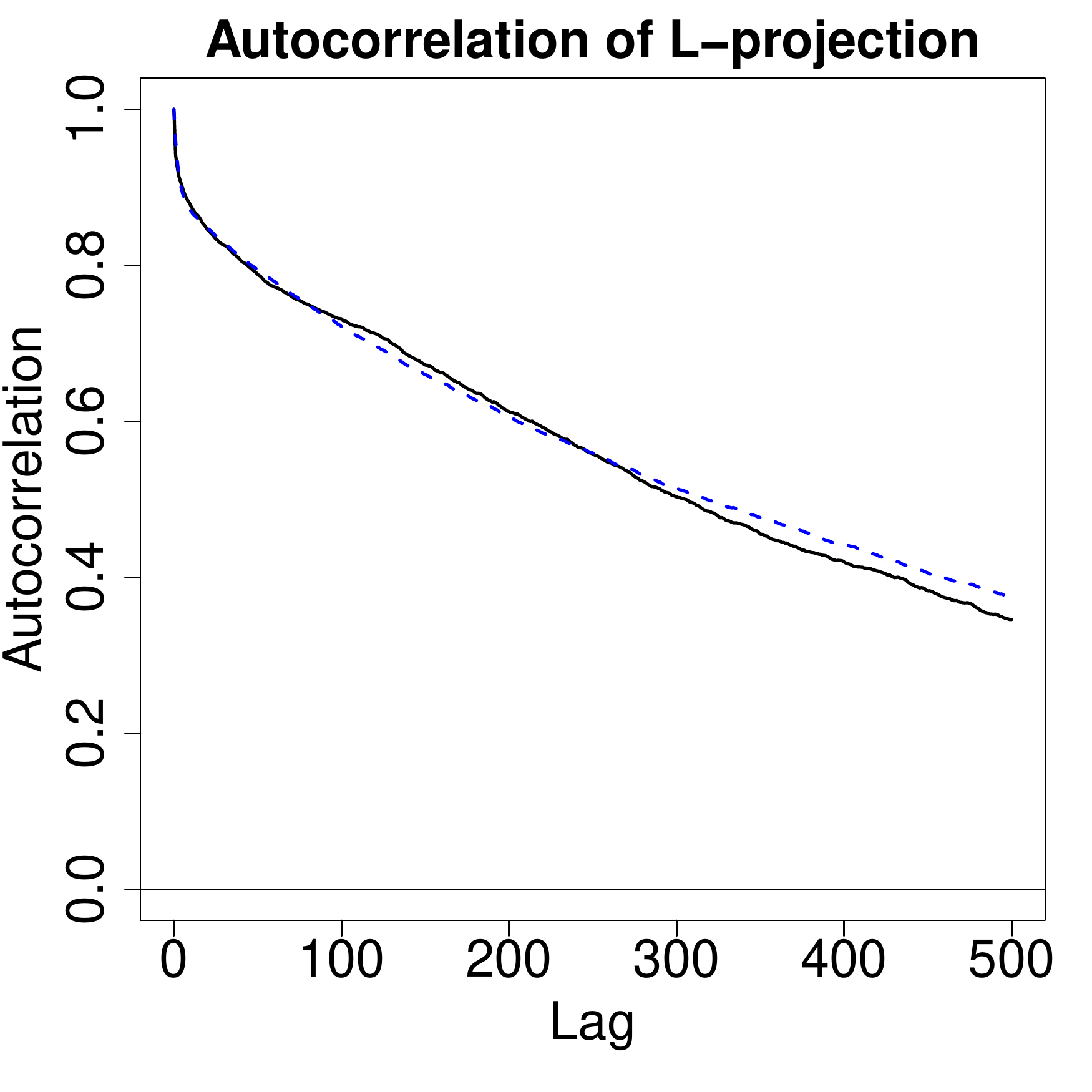} 
\caption{Fit of the diffusion with error model (\ref{e:sumdif}) to the
sample histogram, and sample autocorrelation function of 20.000 observations
of the $L$-reaction coordinate of the small Trp-zipper protein.}
\label{f:simplefit}
\end{center}
\end{figure}

We therefore apply the bimodal diffusion model with additional error (\ref{e:sumdif}) as a second approximation
to the folding dynamics. This agrees well with the empirical autocorrelation function (Figure \ref{f:simplefit}) 
which shows an initial fast drop followed by a long-term slow decay.

Figure \ref{f:simplefit} shows the fit of the pseudo-likelihood function for the marginal bimodal 
normal mixture distribution together with the least squares fit of the autocorrelation function 
(based on the first 100 lags). The parameter estimates are:
\begin{eqnarray*}
\hat{\alpha}=0.55,\quad
\hat{\mu}_1=25.66,\quad
\hat{\mu}_2=30.94,&\quad&
\hat{\sigma}_1=0.54,\quad
\hat{\sigma}_2=1.22,\quad
\hat{\nu}=0.0015,\\
\hat{\gamma}^2=0.88,\quad
&\textnormal{and}&\quad
\hat{\kappa}=0.43.
\end{eqnarray*}
The upper and lower mode of the latent diffusion are estimated by $\hat{u}=\hat{\mu}_1$ 
and $\hat{l}=\hat{\mu}_2$, and the implied folding and unfolding rates of the protein given by 
(\ref{e:oupassage1}) and (\ref{e:oupassage2}) are now estimated by:
\begin{eqnarray*}
\widehat{E}(T_{\ell}^Y|Y_0=u) &=& \frac{\sqrt{2\pi}}{\hat{\nu}}
\int_{\widehat{\tr^{-1}\mu_1}}^{\widehat{\tr^{-1}\mu_2}}\{1-\Phi(x)\}e^{x^2/2}dx
=1138,\\ 
\widehat{E}(T_{u}^Y|Y_0=\ell) &=& \frac{\sqrt{2\pi}}{\hat{\nu}}
\int_{\widehat{\tr^{-1}\mu_1}}^{\widehat{\tr^{-1}\mu_2}}\Phi(x)e^{x^2/2}dx = 1320.
\end{eqnarray*}
Folding and unfolding should thus occur on average once in just over a micro second.
These estimates are realistic for the protein folding in contrary to the ones implied by the 
plain diffusion model without measurement error. 

\begin{figure}[httb]
\begin{center}
\includegraphics[height=5cm,width=7cm]{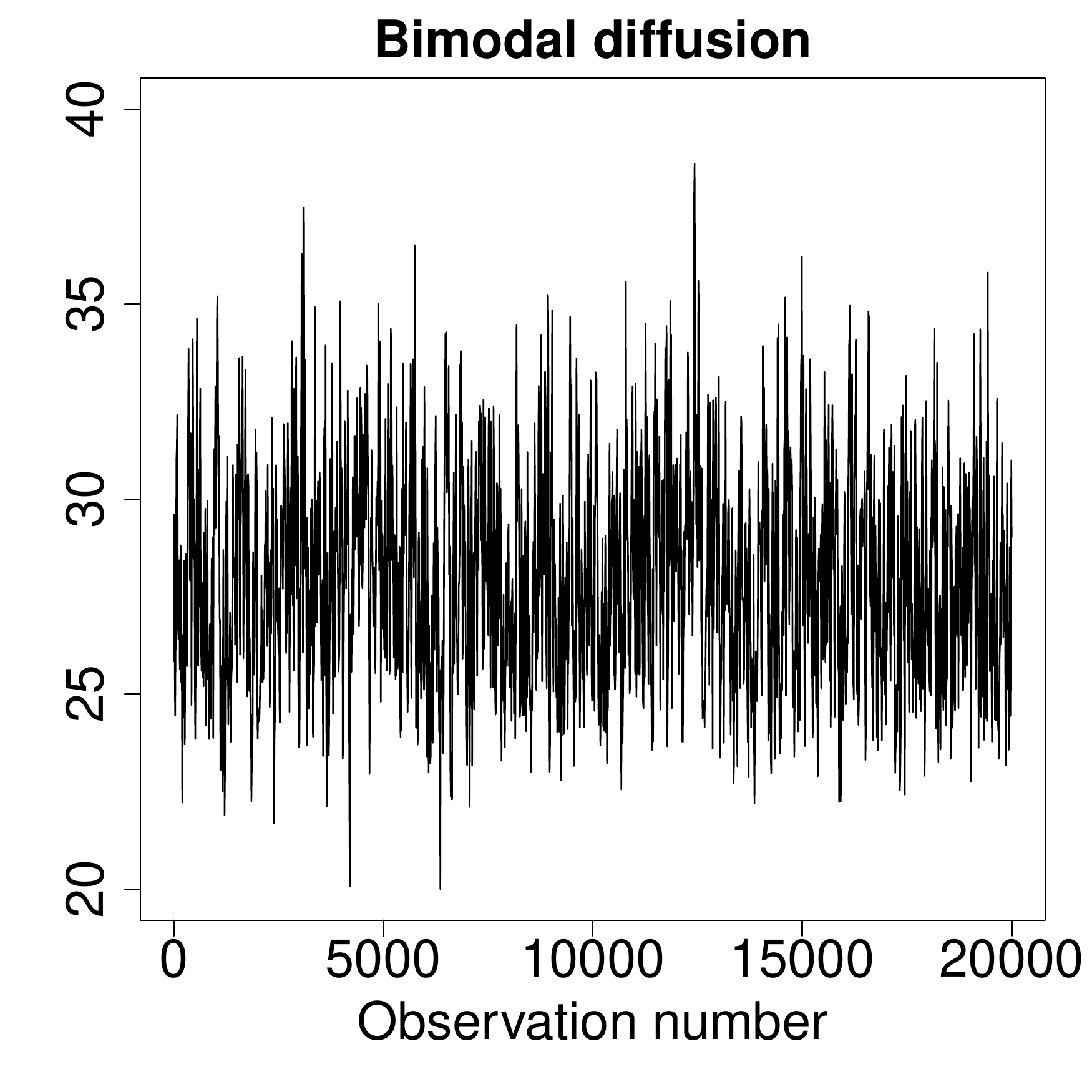} 
\includegraphics[height=5cm,width=7cm]{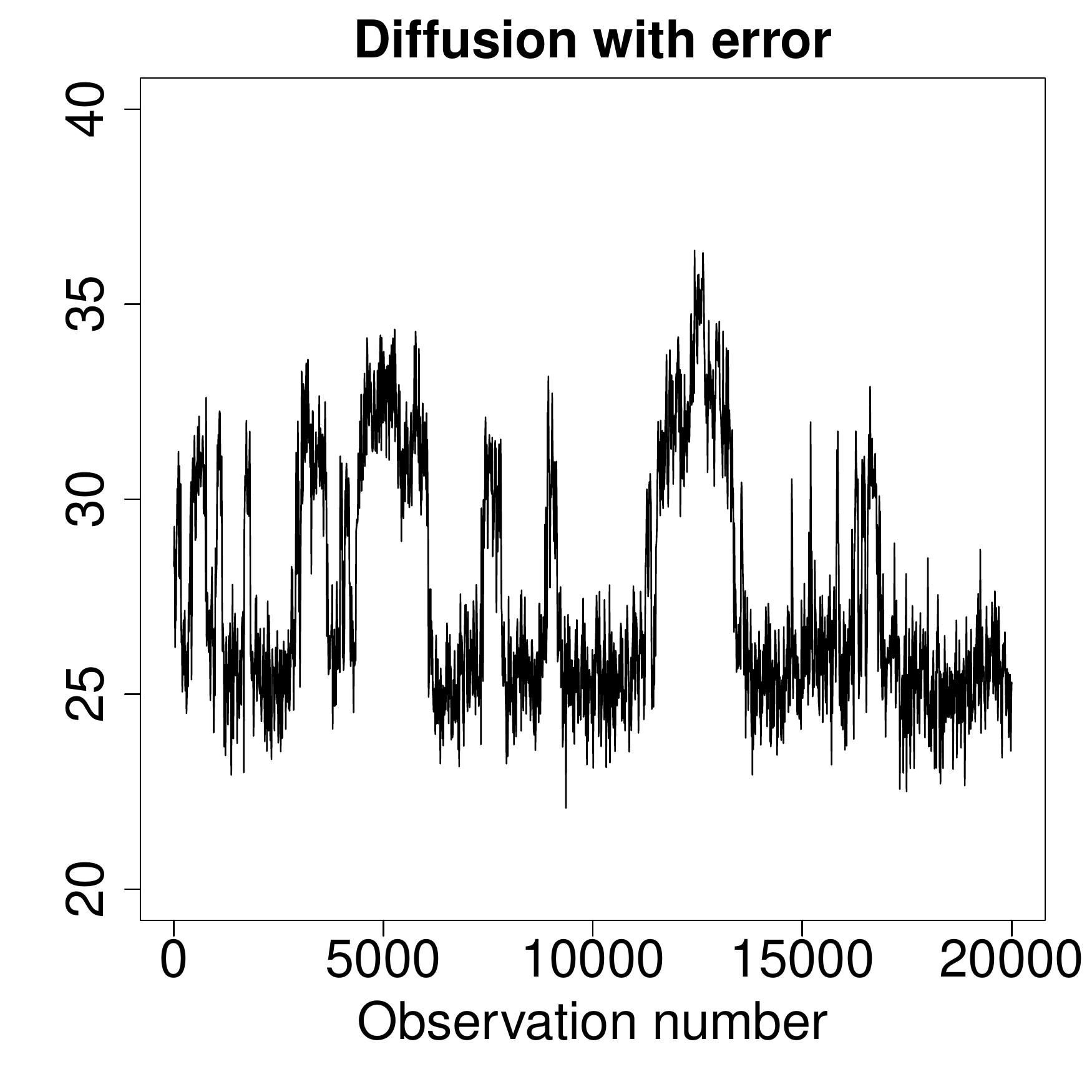} 
\caption{Left panel: A simulated sample path of the bimodal diffusion model (\ref{e:trou}).
Right panel: A simulated sample path of the diffusion plus error model (\ref{e:sumdif}).
Parameters for the simulations were taken from the fit of the two models to the protein data.}
\label{f:simdata}
\end{center}
\end{figure}

As an informal goodness of fit test, we have simulated two trajectories
from the plain diffusion model and the diffusion plus noise model, respectively,
with parameters equal to those estimated for the data. The simulated data are shown 
in Figure \ref{f:simdata}. It is obvious that the model with local error mimics the 
dynamics of the protein data much better than the plain bimodal diffusion model 
(without error) does. 

%The careful reader will note that in spite of its appearant large size the protein dataset
%is in fact not large from an effective sample size perspective due to the strong persistence in
%the autocorrelations. The expected accuracy of the parameter estimates is thus low. Still the data
%is sufficiently large to be used for illustrative purposes and the protein folding rates 
%were estimated to the correct order of magnitude, though only when adjusting for measurement error. 

\section{Conclusion}
\label{s:concl}

Flexible and statistically tractable multi-modal diffusion models have
been developed, where the multi-modality is caused by a combination of
the effects of an energy landscape and of state-dependent random
fluctuations. The new diffusion models were obtained by transformations of
simple well-studied diffusion models. The transformed diffusion was
shown to inherit many properties of the underlying simple diffusion,
including its mixing rates and the distributions of first passage
times. The eigenfunctions of the infinitesimal generator
transform in a straightforward way. Particularly tractable models
are obtained by transformations of the Ornstein-Uhlenbeck
model, but also transformations of  more general Pearson diffusions
give tractable multi-model diffusion models.

Likelihood inference and martingale estimating functions were given
and investigated in the case of a discretely observed bimodal diffusion.
In particular, the theory of martingale estimating functions for
transformed diffusions was generalized to the case of
parameter-dependent transformations. An estimation method was
presented for the case where the bimodal diffusion is observed
with additional correlated measurement error. 

The new approach was applied to molecular dynamics data in form of a reaction coordinate of the
small Trp-zipper protein, where the stationary distribution has two
modes corresponding to a folded and an unfolded state. The folding and
unfolding rates were estimated by the mean passage times between the
two mode points. The new model provides a better fit to this type of
protein folding data than previous models because the diffusion
coefficient is state-dependent, but the fit is far from perfect. A
much more satisfactory fit and realistic folding rates were obtained
by adding correlated measurement error reflecting local features of
the folding funnel of the protein.

\section*{Acknowledgements} 
We are grateful to Sandro Bottero and Jesper Ferkinghoff-Borg, Elektro DTU and Thomas 
Hamelryck, Department of Bioinformatics for supplying the data for the case study and
for introducing us to the interesting problems of modeling and understanding protein
dynamics.

\bibliographystyle{apa}
\bibliography{bimodal_refs}

\end{document}